\documentclass[aps,twocolumn,pra,preprintnumbers,superscriptaddress,amsmath,amssymb]{revtex4-1}
\usepackage{graphicx}
\usepackage{color, colortbl}
\definecolor{Gray}{gray}{0.85}
\usepackage{dcolumn}
\usepackage{bm}
\usepackage{tikz} \usepackage{mathtools}
\usepackage{tabularx}
\newcommand*\diff{\mathop{}\!\mathrm{d}}
\usepackage{braket} \usepackage[caption=false]{subfig} \usepackage{natbib}
 \usepackage{hyperref}
\hypersetup{ colorlinks, allcolors=[rgb]{.18,.19,.57} }
 
\usepackage[textsize=tiny]{todonotes} \usepackage{setspace} \makeatletter
\renewcommand{\todo}[2][]{%
  \@todo[caption={#2}, #1]{\begin{spacing}{0.5}#2\end{spacing}}%
}\makeatother

\begin{document}
\title{Bi-selective pulses for large-area atom
  interferometry}
\author{Jack Saywell} \email[]{j.c.saywell@soton.ac.uk} \author{Max Carey}
\affiliation{School of Physics \& Astronomy, University of Southampton,
  Highfield, Southampton, SO17 1BJ, UK} \author{Ilya Kuprov} \affiliation{School
  of Chemistry, University of Southampton, Highfield, Southampton, SO17 1BJ, UK}
\author{Tim Freegarde} \affiliation{School of Physics \& Astronomy, University
  of Southampton, Highfield, Southampton, SO17 1BJ, UK} \date{\today}

\begin{abstract}
We present designs for the augmentation ‘mirror’ pulses of large-momentum-transfer atom interferometers that maintain their fidelity as the wavepacket momentum difference is increased. These bi-selective pulses, tailored using optimal control methods to the evolving bi-modal momentum distribution, should allow greater interferometer areas and hence increased inertial measurement sensitivity, without requiring elevated Rabi frequencies or extended frequency chirps. Using an experimentally validated model, we have simulated the application of our pulse designs to large-momentum-transfer atom interferometry using stimulated Raman transitions in a laser-cooled atomic sample of $^{85}$Rb at 1 $\mu$K. After the wavepackets have separated by 42 photon recoil momenta, our pulses maintain a fringe contrast of 90\% whereas, for adiabatic rapid passage and conventional $\pi$ pulses, the contrast is less than 10\%. Furthermore, we show how these pulses may be adapted to suppress the detrimental off-resonant excitation that limits other broadband pulse schemes.
\end{abstract}

\maketitle

\section{\label{sec:Introduction}Introduction}
Atom interferometers \cite{RBerman1997} reverse the optical interferometry roles of light and matter by using pulses of laser light to split, redirect, recombine and interfere atomic matter waves, allowing the precise measurement of gravitational fields \cite{Peters2001, Mcguirk2002, Rosi2014}, inertial motions \cite{Gustavson1997, Barrett2014, Hoth2016} and more esoteric fields \cite{Hamilton2015a, Jaffe2017} to which optical interferometers would have little or no sensitivity.

Most atom interferometers for inertial sensing use Bragg \cite{Muller2008c} or Raman \cite{Kasevich1991a,Kasevich1991b} transitions driven by counter-propagating laser pulses as the beamsplitters and mirrors that split and direct the atomic wavepackets. The two-photon recoil accompanying these stimulated scattering processes imparts a momentum difference of 2$\hbar$k between the two interferometer paths, where k is the single-photon wave number. As with an optical interferometer, the measurement sensitivity depends upon the spatial area enclosed. This is proportional to the momentum separation, and can hence be increased by using additional mirror pulses - known as augmentation pulses - to impart further impulses to the atomic wavepackets forming a large momentum transfer (LMT) interferometer \cite{McGuirk2000}. 

In practice, inhomogeneities such as variations in beam intensity and atomic velocity reduce the fidelity of the augmentation pulses. The accrued effect of such imperfections limits the fringe visibility and can reverse the LMT sensitivity gains \cite{McGuirk2000, Butts2013, Kotru2015}. Cooling the atoms or filtering their velocity distribution can improve the initial fidelity at the expense of a weaker signal, but a velocity spread is inevitable in LMT interferometry because of the increasing momentum separation introduced between the primary interfering arms. Techniques such as increasing the Rabi frequency \cite{Rudolph2020} and employing beam shaping optics \cite{Mielec2018} increase the experimental complexity and power requirements of the interferometer.

An alternative approach is to adopt robust atom-optics that retain high fringe visibility for large momentum separations. Composite \cite{Levitt1981a, Dunning2014} and adiabatic
rapid passage (ARP) \cite{Bateman2007} Raman pulses have been investigated,
achieving momentum splittings of 18$\hbar$k \cite{Butts2013} and 30$\hbar$k
\cite{Kotru2015} respectively. Both these techniques can increase the velocity
acceptance of the augmentation pulses at the expense of an increase in pulse
duration, but in practice fringe visibility was lost after 4-7 augmentation pulses. Alternative atom-optics to Raman transitions such as multiphoton Bragg pulses \cite{Muller2008c, Chiow2011} and Bloch oscillations within optical lattices \cite{Clade2009,Muller2009,McDonald2013}  have achieved impressive momentum transfer, but these put even more stringent demands on the initial momentum
distribution \cite{Szigeti2012} and, when used for interferometry, similarly
limit the signal-to-noise ratio (SNR) \cite{Butts2013, Kotru2015}.

We have previously used optimal control techniques \cite{Khaneja2005, DeFouquieres2011} to design robust high fidelity pulses for small momentum transfer interferometry \cite{Saywell2018b, Saywell2020a}. In this paper, we address LMT, by designing ``bi-selective'' pulses that offer high fidelity for two ranges of velocity that track the atoms in the two arms of the LMT interferometer, allowing high fringe contrast to be maintained for larger interferometer areas. We compare our pulses with conventional composite pulses and ARP through simulations using an previously validated approach \cite{Dunning2014, Saywell2020a, Kotru2015}, and consider an extension to reduce unwanted double diffraction \cite{Leveque2009, Malossi2010a}. Our method is a departure from previous approaches in which the same robust pulse has been used for every augmentation pulse in the interferometer
sequence, ultimately limiting the achievable momentum splitting to the velocity
acceptance of the pulse. The technique has general applicability because many interferometer arrangements are already set up to implement similar modulation sequences and, as the algorithm optimises tolerance of variations, the designs do not depend critically upon experimental parameters.

\section{LMT atom interferometers}

\begin{figure*}[!htb]
  \centering \includegraphics[width=\linewidth]{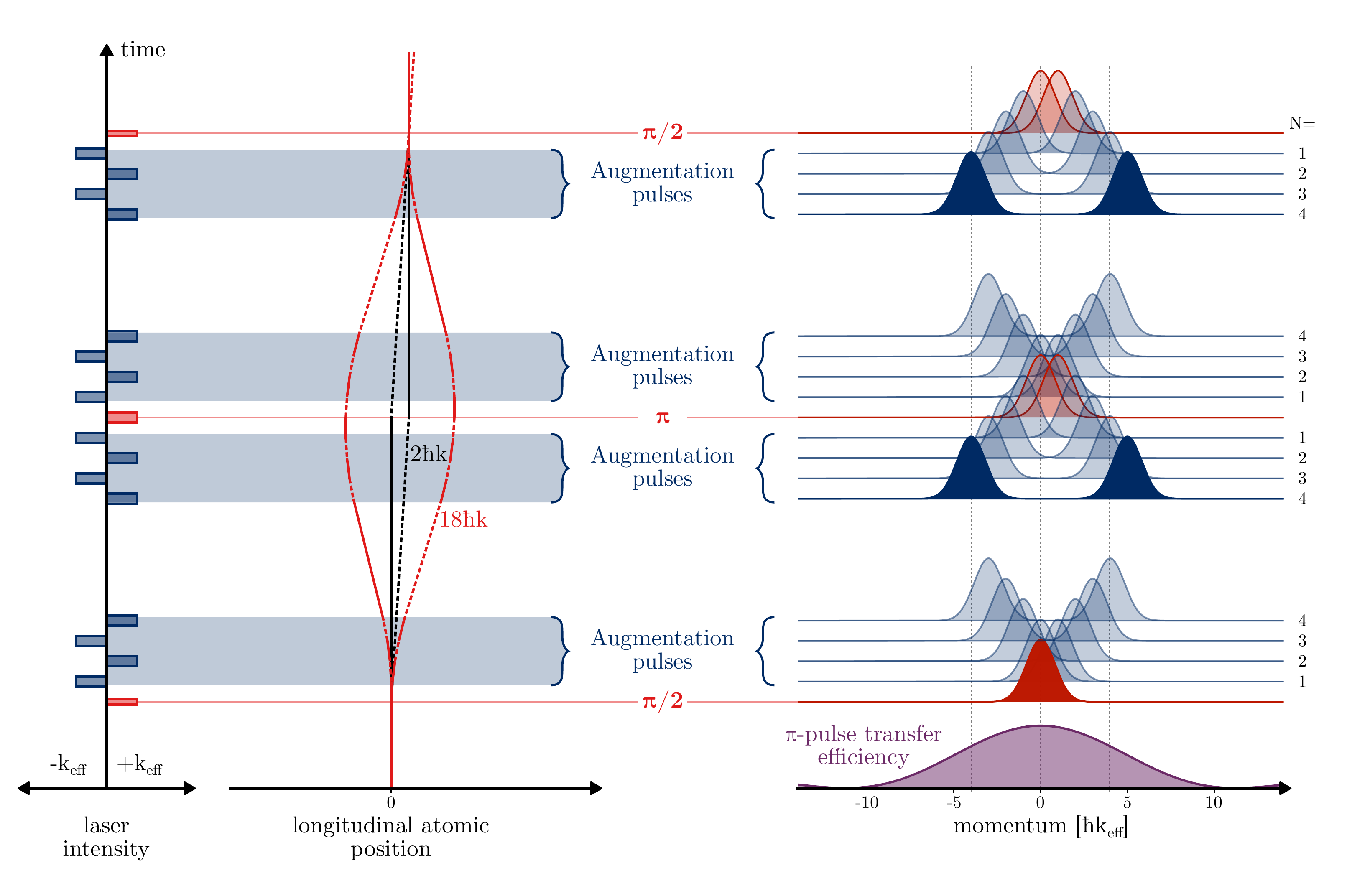}
  \caption{Atomic state trajectories (left) as a function of position within an LMT interferometer sequence showing the case where the beamsplitter and mirror operations are extended by a sequence of 4 augmentation pulses (18$\hbar$k trajectory). The initial momentum distribution (right), represented by the red shaded region, is separated into two arms. The momentum distribution seen by each pulse throughout the LMT sequence for $^{85}$Rb atoms with a Maxwell-Boltzmann temperature of $\sim 1~\mu$K is shown by the blue shaded regions on the right. As more pulses are added to increase the interferometer area, the separation in the resonance conditions for the arms begins to exceed the velocity acceptance of a $\pi$-pulse, shown bottom right for a Rabi frequency of 200 kHz. }
  \label{fig:LMT_diagram}
\end{figure*}

Figure \ref{fig:LMT_diagram} shows the space-time diagram of a typical Raman
LMT interferometer. While the standard $\pi/2-\pi-\pi/2$ interferometer sequence
produces a momentum splitting between the interferometer arms of $\hbar
\mathrm{k}_{\mathrm{eff}}$, where ${k}_{\mathrm{eff}} = 2$k is the effective
wavevector of the atom-optics, a greater momentum splitting can be achieved by augmenting the mirror and beamsplitter pulses with additional pulses with alternating wavevectors \cite{McGuirk2000}. These augmentation pulses
are designed to transfer atomic population between two internal states and
increase the momentum splitting between the wavepackets in the interferometer.
By extending the beamsplitter operation by $\mathrm{N}$ augmentation pulses, the
momentum splitting may be increased to $(2\mathrm{N} + 1)\hbar
\mathrm{k}_{\mathrm{eff}}$, increasing the intrinsic phase sensitivity of
the interferometer.

However, this increased splitting also introduces differential Doppler shifts
between the interfering arms, quantised in multiples of the two-photon recoil
shift $\delta_{\mathrm{recoil}} = \hbar \mathrm{k}_{\mathrm{eff}}^2 /2m$, that
depend on the atomic mass $m$. Indeed, this is automatic, since the Doppler shift and inertial sensitivity have the same origin. For the $n$th augmentation pulse
($n=1,2,3,...,\mathrm{N}$), the arms of the interferometer
have Raman resonance conditions separated by 4$n\delta_{\mathrm{recoil}}$
\cite{Butts2013, Kotru2016a, Jaffe2019} (for Raman transitions on the $^{85}$Rb
D2 line, $\delta_{\mathrm{recoil}} \approx 2\pi \times 15.4$ kHz). The distribution of
detunings for the $n$th augmentation pulse may thus be visualised as two
Gaussian distributions, corresponding to the Maxwell-Boltzmann temperature of
the atomic source, separated in frequency space by 4$n\delta_{\mathrm{recoil}}$. As illustrated in Figure \ref{fig:LMT_diagram}, the augmentation pulses are
only efficient as long as this split distribution fits within the velocity acceptance of the pulses. For conventional $\pi$-pulses this can span just a few recoil momenta, fundamentally limiting the momentum transfer achievable before fringe visibility is lost \cite{McGuirk2000, Butts2013}.

Many composite and shaped pulses have been developed in the field of nuclear magnetic resonance (NMR) spectroscopy to improve the control of nuclear spins in the presence of experimental inhomogeneities. Composite pulses \cite{Levitt1981a} replace single rectangular pulses with sequences of pulses of varying durations and phases. The overall effect
of the pulse sequence is to replicate the operation of a single pulse, but with
an increased tolerance of unwanted variations in coupling strength, detuning
from resonance, or both. Dunning \textit{et al.} \cite{Dunning2014} investigated the potential fidelity
improvements various established composite pulses might bring in atom interferometry compared
with rectangular $\pi$ pulses. Of the pulses tested, the WALTZ pulse
\cite{Shaka1983}, a relatively short three step robust state-transfer pulse, had
the highest fidelity and velocity acceptance. Similarly, Butts \textit{et al.}
\cite{Butts2013} employed WALTZ to improve the contrast in an LMT
interferometer, doubling the sensitivity.

An alternative class of robust pulses is known as frequency
swept adiabatic rapid passage (ARP). ARP can transfer atomic population between the states of a two-level atom with high efficiency \cite{Bateman2007, Kovachy2012a, Kotru2015, Jaffe2018}. During an ARP pulse, the driving field frequency is swept slowly through resonance such that the atomic state follows the evolution of the field and may be moved with high precision anywhere on the surface of the Bloch sphere \cite{Feynman1957}. In this picture, the quantum state vector precesses about an instantaneous rotation axis, the field vector, defined by the amplitude, detuning, and phase of the driving field. For ARP to be efficient, the pulse must satisfy a condition of adiabaticity: the motion of the field vector on the Bloch sphere must be slower than the rate at which the atomic state precesses about it \cite{Bateman2007}. The simplest example of ARP is the linear frequency chirp, where the pulse amplitude, or Rabi frequency, is fixed, and the laser detuning is swept linearly through resonance. This pulse can be efficient and robust, but it is necessarily long in comparison to
rectangular and composite pulses, thereby increasing the risk that coherence is
lost through spontaneous emission.

By allowing the pulse amplitude to vary, various schemes have been developed which
outperform the linear frequency chirp \cite{Hardy1986, Baum1985b, Tannus1996,
Hwang1998, Garwood2001a}. These pulses have high efficiency and robustness,
maintaining adiabaticity while reducing pulse duration. An example is the Tanh/Tan ARP pulse \cite{Hwang1998}, where the frequency
sweep follows a tangential function of time and the pulse amplitude follows a
hyperbolic tangential function of time. The Tanh/Tan pulse was implemented in a large-area atom interferometer by Kotru \textit{et al.}
\cite{Kotru2015} increasing the interferometer contrast over that obtained using
simple $\pi$ pulses and achieving a momentum splitting of 30 $\hbar$k in a 9
$\mu$K $^{133}$Cs atomic sample.

Although ARP can obtain impressive population transfer efficiency, with a
detuning bandwidth that increases with the pulse duration, its potential utility
in interferometry is limited by the dynamic phase imprinted on the diffracting
wavepackets and the variation in effective time origin \cite{Bateman2007, Kotru2016a, Jaffe2018, Jaffe2019}. The dynamic phase depends on the optical intensity, and rapid dephasing is therefore inevitable when the atom cloud expands through variations in laser intensity. In practice this effect is
limited because ARP pulses applied in quick succession approximately cancel the
dynamic phase, but it leads to a trade-off between longer and theoretically more
efficient pulses, and dephasing caused by imperfect dynamic phase cancellation
when the beam quality is non-ideal \cite{Kotru2015}. There is therefore a need
for pulses which match or improve upon ARP in terms of state transfer
efficiency, but which are robust to temporal variations in the Rabi frequency
during the interferometer sequence.

LMT interferometry requires augmentation pulses that provide efficient population transfer across the atom cloud and throughout the sequence. To maintain sensitivity and prevent loss of fringe visibility, the interferometer sequence itself should impart the same phase to all atoms, and all atoms should have the same sensitivity to the external influence being measured. Pulses need not satisfy these conditions individually, provided that subsequent cancellation achieves them for the sequence as a whole. ARP pulses, for example, have detuning-dependent effective time origins and dynamic phases which can be cancelled by a later pulse provided there is no variation in optical intensity.

\section{Optimal control techniques}
Robust pulses may be generated with optimal control techniques
\cite{Kobzar2004, Kobzar2012} by dividing a pulse into discrete time slices and
treating the phase and/or amplitude of each slice as control parameters that may
be adjusted to optimise the fidelity of a desired optimisation.

In a typical Raman atom interferometer, two-photon transitions
are driven by lasers far from single-photon resonance so that the intermediate
level can be adiabatically eliminated to leave an effective two-level system
between stable states $\ket{\mathrm{g}}$ and $\ket{\mathrm{e}}$. The atom optics then effect a
rotation of the quantum state on the surface of the Bloch sphere for this basis
at a rate $\tilde{\Omega}_R = \sqrt{\Omega_R^2 + \delta^2}$, about an axis (the
field vector)
\begin{equation}\label{eq: field_vector}
  \mathbf{\Omega} = \Omega_R\cos(\phi_L)\hat{x} + \Omega_R\sin(\phi_L)\hat{y} +\delta\hat{z},
\end{equation}
determined by the relative laser phase $\phi_L$, the two-photon Rabi frequency on
resonance $\Omega_R$, and the Raman detuning
\begin{equation}
  \delta = (\omega_1-\omega_2)  - \omega_{\mathrm{eg}} + \delta_{\mathrm{Doppler}} + \delta_{\mathrm{recoil}}
\end{equation}
that includes terms for the two-photon recoil shift $\delta_{\mathrm{recoil}} =
\hbar \mathrm{k}_{\mathrm{eff}}^2 /2m$ and the Doppler shift
$\delta_{\mathrm{Doppler}} = \mathbf{k}_{\mathrm{eff}}\cdot \mathbf{p}/m$, which
depends on the initial momentum $\mathbf{p}$ of each atom in the interferometer. $\omega_{1,2}$ are the laser frequencies and $\omega_{\mathrm{eg}}$ is the frequency splitting of the levels $\ket{\mathrm{g}}$ and $\ket{\mathrm{e}}$.
This rotation can be written in terms of a propagator \begin{equation} \label{eq: prop} \hat{U} = \begin{pmatrix}
    C^* & -iS^* \\
    -iS & C
  \end{pmatrix}
\end{equation} acting on the ($\ket{g}$, $\ket{e}$) basis, where $C$ and $S$ are the
`continuing' and `scattering' amplitudes defined as \cite{Stoner2011}
\begin{flalign} \label{eq: propagator_elements} \nonumber
  C &\equiv \cos\big({\tilde{\Omega}_R\tau }\big/{2}\big)+ i\big({\delta}\big/{ \tilde{\Omega}_R}\big)\sin\big({\tilde{\Omega}_R\tau}\big/{2}\big)\\
  S &\equiv e^{i\phi_L}\big({{\Omega}_R }\big/{
    \tilde{\Omega}_R}\big)\sin\big({\tilde{\Omega}_R\tau }\big/{2}\big).
\end{flalign}

The conventional beamsplitter and mirror pulses of atom interferometry have a
fixed amplitude and phase, and the durations of the interactions are set so
that, on resonance, they perform $\pi/2$ and $\pi$ rotations respectively.
Composite and ARP pulses depend on varying the pulse parameters $\Omega_R(t)$,
$\phi_L(t)$, and $\delta(t)$ as a functions of time $t$ so that, for a given atom, the rotation axis
and rate change during the pulse.

The action of such a pulse on the quantum state can be evaluated efficiently by dividing the pulse into discrete timesteps of duration $\diff t$, with the propagator for an entire
pulse given by the time-ordered product of propagators for each slice,
calculated according to Equation \eqref{eq: prop}. A pulse is then described by
piece-wise constant waveforms $\Omega_R(t)$, $\phi_L(t)$, and $\delta(t)$ that
constitute a finite number of control parameters.  Upon defining a
suitable fidelity, such as a measure of how accurately a given initial state
$\ket{\psi(t=0)}$ is driven to a target $\ket{\psi_{\mathrm{T}}}$ by the pulse, optimal waveforms may be found using numerical routines from optimal control theory. In this work we focus on producing pulses with
optimised phase and amplitude profiles $\phi_L(t)$ and $\Omega_R(t)$.

One efficient and popular quantum control algorithm, originally developed for
NMR applications, is gradient ascent pulse engineering, or GRAPE
\cite{Khaneja2005}. GRAPE computes derivatives of the pulse fidelity with
respect to the control parameters without the need for computationally expensive
finite-differencing methods, and can be used in conjunction with the
limited-memory Broyden-Fletcher-Goldfarb-Shanno (L-BFGS) quasi-Newton method
\cite{DeFouquieres2011}. By computing the gradient and estimating
the Hessian of the fidelity landscape, and provided local maxima with low-fidelity
are avoided, the algorithm can find an optimum solution efficiently.

We consider the following two possible fidelities for augmentation pulses,
\begin{flalign} \label{eq:fidelities1} \mathcal{F}_{\mathrm{square}} & =
  |\braket{\mathrm{e}|\hat{U}|\mathrm{g}}|^2 \\ \label{eq:fidelities2}
  \mathcal{F}_{\mathrm{real}} & =
  \mathrm{Re}(\braket{\mathrm{e}|\hat{U}|\mathrm{g}}).
\end{flalign}
Both of these fidelities, if maximised, will yield pulses that efficiently
transfer population from one basis state to the other, the essential requirement
of an augmentation pulse. If the fidelities are averaged over an ensemble of
detunings and a range of Rabi frequencies, the resulting pulse will be made
robust to these specific errors. Maximising $\mathcal{F}_{\mathrm{square}}$
leads to pulses where the phase of the overlap
$\braket{\mathrm{e}|\hat{U}|\mathrm{g}}$ is unimportant but the quantum state is
rotated by 180$^{\circ}$ from pole-to-pole on the Bloch sphere. Conversely,
maximising $\mathcal{F}_{\mathrm{real}}$ leads to pulses where the phase of the
overlap $\braket{\mathrm{e}|\hat{U}|\mathrm{g}}$ is well-defined.

If the phase of the overlap $\braket{\mathrm{e}|\hat{U}|\mathrm{g}}$ varies from
pulse-to-pulse across an atomic sample in an LMT sequence, the resulting
interference fringes can be washed out when the contributions from each atom in
the cloud are averaged at the end of the interferometer. Providing the Rabi
frequency does not change much between augmentation pulses, these phases
approximately cancel in LMT interferometers but if the Rabi frequency varies
temporally throughout the interferometer, rapid dephasing can occur. This is
readily observed with ARP, which requires a high degree of cancellation in the
phase factors introduced by each pulse in the sequence \cite{Kotru2015, Jaffe2018, Jaffe2019}. Optimising
$\mathcal{F}_{\mathrm{real}}$ for a range of Rabi frequencies will mean the
phase of the overlap $\braket{\mathrm{e}|\hat{U}|\mathrm{g}}$ remains fixed even
if the coupling strength varies within that range. Therefore, we expect
$\mathcal{F}_{\mathrm{real}}$ to yield augmentation pulses that are insensitive
to dephasing caused by temporal changes in the Rabi frequency due, for example,
to noisy beam intensity profiles and/or the ballistic expansion of the atom
cloud.

\section{Bi-selective Raman pulses}
For the nominal three-pulse $\pi/2 -\pi -\pi/2$ interferometer, the Raman detunings arise from the velocity components parallel to the Raman beam axis and hence the temperature of the atomic cloud. As the
sequence is extended by augmentation pulses with alternating effective
wavevector directions, the resonant frequencies of each interferometer arm
separate due to the imparted photon momentum (shown on the right-hand side of Figure \ref{fig:LMT_diagram}). This means the detuning distribution splits in two, and continues to separate as more momentum is
imparted.  Optimal control can optimise a pulse for any assumed velocity distribution \cite{Kobzar2005}. We use this feature to design “bi-selective” pulses, tailored to yield efficient population transfer only in the frequency ranges occupied by the primary interferometer arms. By optimising each pulse in the sequence individually, as opposed to employing a single augmentation pulse design throughout the interferometer  \cite{McGuirk2000, Butts2013, Kotru2015}, we can therefore take advantage of the fact that their transfer efficiency in the frequency space in between the two arms is unimportant. This allows bi-selective pulses to be shorter than ARP, composite or optimised shaped pulses of the equivalent efficiency.

We express the bi-selective fidelity for the $n$th augmentation pulse as
\begin{equation}\label{eq: bi-selective fidelity}
  \mathcal{F}^{\mathrm{A}}_n = \sum_{\delta,\Omega_R \in\mathrm{L}_n} \mathcal{F}(\delta,\Omega_R) +  \sum_{\delta,\Omega_R \in \mathrm{U}_n} \mathcal{F}(\delta,\Omega_R),
\end{equation}
where $\mathcal{F} = \mathcal{F}_{\mathrm{real},\ \mathrm{square}}$ is the
single-atom augmentation pulse fidelity (Equations \ref{eq:fidelities1} and
\ref{eq:fidelities2}) and $\mathrm{L}_n$ and $\mathrm{U}_n$ are the ensembles
representing the atomic frequency distributions of the lower and upper arms
during the $n$th augmentation pulse respectively. In order to normalise the fidelity such that the maximum value is unity, we divide Equation \ref{eq: bi-selective fidelity} by the number of detunings and amplitude errors included in the entire ensemble.

We compose the detuning ensembles for the lower and upper interferometer arms for
the $n$th augmentation pulse from two uniform discrete distributions centred at
$\pm 2n \delta_{\mathrm{recoil}}$. The ensembles used in the optimisation should represent the velocity distribution of the atoms. However, in order to reduce computation time we approximate the true distribution using two uniform distributions each with a sample size of 20. The range spanned by each distribution is given by 4 standard
deviations of detuning arising from the velocity distribution along the Raman
beam axis, which we assume follows a Maxwell Boltzmann distribution.

Our LMT pulses are individually optimised, which assumes there are no correlations in residual errors between pulses. However, It may be possible to develop alternative measures of performance which reflect the fidelity of the entire interferometer and allow pulses within a sequence to be optimised cooperatively \cite{Braun2014}, for example by compensating each other's imperfections. 

An alternative approach to achieve bi-selectivity involves concatenating two inversion pulses with the appropriate frequency shifts to address each interferometer arm separately. While this makes the individual optimisations simpler, the resulting pulses are longer, give different velocity classes different pulse origins, and lose fidelity as the classes began to overlap. A simple superposition of the two components would cure the problems of length and origin, though not overlap, and requires amplitude modulation in addition \cite{Emsley1990}.

The concept of bi-selective pulses for LMT interferometry bears resemblance to
the technique of band-selective pulses in NMR spectroscopy
\cite{Emsley1990, Geen1991, Freeman1998a, Janich2011}, where pulses have been designed to excite or invert nuclear spins within single or multiple frequency bands, but suppress the
response of spins outside the desired frequency range. Typically, these pulses require smooth waveforms because the response at large resonance offsets is determined by the Fourier transform of the pulse shape \cite{Warren1984}. As a result, composite pulses, which are composed of concatenated sequences of constant amplitude pulses with discrete phase shifts, lead to non-negligible excitation far off-resonance.

The suppression of the action of a pulse outside a specific frequency range may be useful in interferometer geometries where, for example, the counter-propagating Raman beams are obtained by retroreflection of a single beam with both Raman frequencies and there are necessarily four frequency components that may interact with the atom cloud. When there is no acceleration of the atoms along the beam axis then double-diffraction interferometry schemes can be be employed that make use of all of the frequency components simultaneously  \cite{Leveque2009, Malossi2010a} . However, in vertically-orientated, ground-based atom interferometers, the Doppler shift caused by gravitational acceleration is commonly used to isolate a single frequency pair by shifting the other pair off resonance \cite{Butts2013}. However, broadband pulses such as composite or ARP pulses often have non-zero transfer efficiency at large detunings \cite{Kotru2016a, Warren1984}, meaning one must wait longer to ensure negligible excitations from the off-resonant pair, costing time which may otherwise be used to enhance the sensitivity.

By directly suppressing the transfer efficiency outside a specific
range of frequencies, such off-resonant excitation with broadband pulses may be
avoided. We achieve this suppression of unwanted excitation by modifying our
bi-selective pulse fidelity (Equation \ref{eq: bi-selective fidelity}), adding the following
penalty term which represents the suppression band of detunings where the transfer
efficiency should be minimised,
\begin{equation}\label{eq: bi-selective fidelity suppress}
  \mathcal{F}_\mathrm{suppress} = \sum_{\Omega_R, \delta = \delta_{\mathrm{min}}}^{\delta_{\mathrm{max}}} |\braket{\mathrm{g}|\hat{U}|\mathrm{g}}|^2 + \sum_{\Omega_R, \delta = -\delta_{\mathrm{min}}}^{-\delta_{\mathrm{max}}} |\braket{\mathrm{g}|\hat{U}|\mathrm{g}}|^2.
\end{equation}
Here, $\delta_{\mathrm{min},\mathrm{max}}$ represent the initial and final Raman
detuning values for the two suppression bands.

\section{Results}

\begin{figure*}[!htb]
  \centering \includegraphics[width=\linewidth]{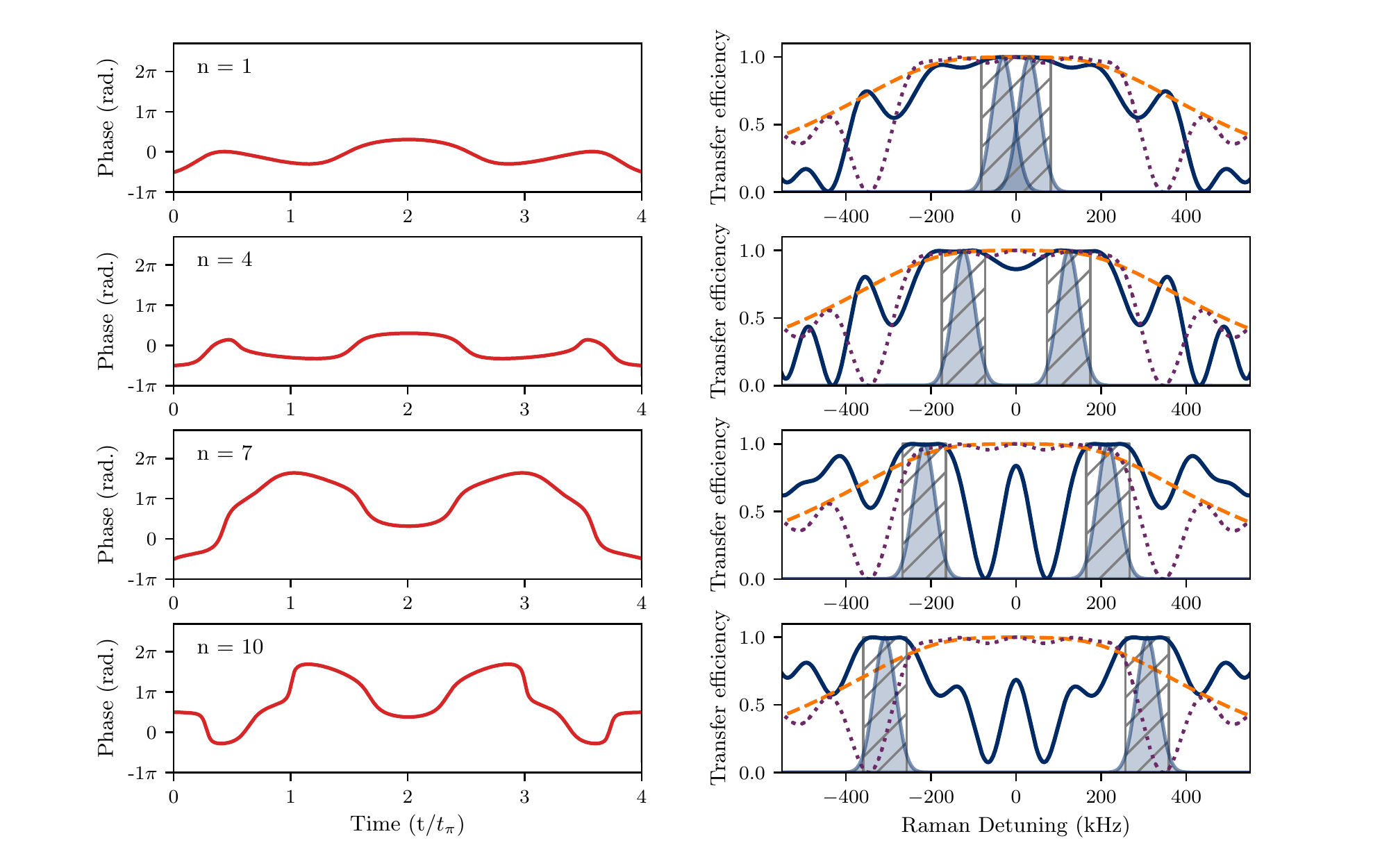}
  \caption{Bi-selective phase-modulated augmentation pulses optimised with the phase-sensitive fidelity $\mathcal{F}_{\mathrm{real}}$ for a temperature of 1 $\mu$K. The resulting phase profiles for the
    1\textsuperscript{st}, 4\textsuperscript{th}, 7\textsuperscript{th}, and
    10\textsuperscript{th} augmentation pulses are shown in the left panel. The
    right panel shows the simulated transfer efficiency of each bi-selective
    pulse (blue solid curve), the ARP Tanh/Tan pulse of equivalent length
    (orange dashed curve), and the WALTZ composite pulse (purple dotted curve) as a function of the Raman detuning. The corresponding detuning distributions seen by each augmentation pulse for the primary interferometer arms are shown by the shaded regions. The range of detunings included in each optimisation is shown by the hatched regions.}
  \label{fig:bi-selective-waveforms}
\end{figure*}

We have optimised phase-modulated, constant amplitude, bi-selective augmentation
pulses with GRAPE, using both the phase insensitive fidelity
$\mathcal{F}_{\mathrm{square}}$ and the phase sensitive fidelity
$\mathcal{F}_{\mathrm{real}}$. We have investigated their performance through
simulation of the resulting interferometer contrast in a laser-cooled sample of
$^{85}$Rb at 1 $\mu$K. In all optimisations, the timestep of the
pulses was 25 $n$s, and the effective Rabi frequency was 200 kHz,
meaning the duration of a rectangular $\pi$ pulse ($t_{\pi}$) should be 2.5
$\mu$s in the absence of any detuning or amplitude error. The pulses were
optimised for an atomic temperature of 1 $\mu$K, and a range of amplitude errors
of $\pm$10 \% the effective Rabi frequency. The length of the augmentation pulses was fixed to be 4$t_{\pi}$. This duration allowed for a sufficiently good pulse to be obtained when using both fidelities, although optimising $\mathcal{F}_{\mathrm{real}}$ typically requires a longer pulse to reach an equivalent fidelity to the phase-insensitive case. The initial guess for the phase profile of the pulses was in each case a
sequence of random phases. We keep the pulse amplitude constant when optimising the bi-selective fidelity  $\mathcal{F}^{\mathrm{A}}_n$ (Equation \ref{eq: bi-selective fidelity}) because allowing the amplitude to vary does not lead to a higher terminal fidelity in this case.

The NMR spin simulation software suite for MATLAB, \textit{Spinach}
\cite{Hogben2011}, and its optimal control module, were modified to optimise
bi-selective pulses using the L-BFGS GRAPE method \cite{DeFouquieres2011}. Each pulse optimisation was
set to terminate following 300 iterations or when either the norm of the fidelity gradient became smaller than
10$^{-7}$ or the norm of the step size dropped below 10$^{-3}$. The resulting waveforms, optimising the fidelity $\mathcal{F}_{\mathrm{real}}$, are shown in Figure \ref{fig:bi-selective-waveforms}, for pulses tailored to the velocity distributions expected for the 1\textsuperscript{st}, 4\textsuperscript{th}, 7\textsuperscript{th}, and
10\textsuperscript{th} augmentation pulses of the LMT sequence. Efficient population transfer is achieved in each case. Interestingly, smooth and symmetrical waveforms are found despite there being no constraint on symmetry or waveform smoothness included in the
optimisation. 

\begin{figure}[!htb]
  \centering \includegraphics[width=\columnwidth]{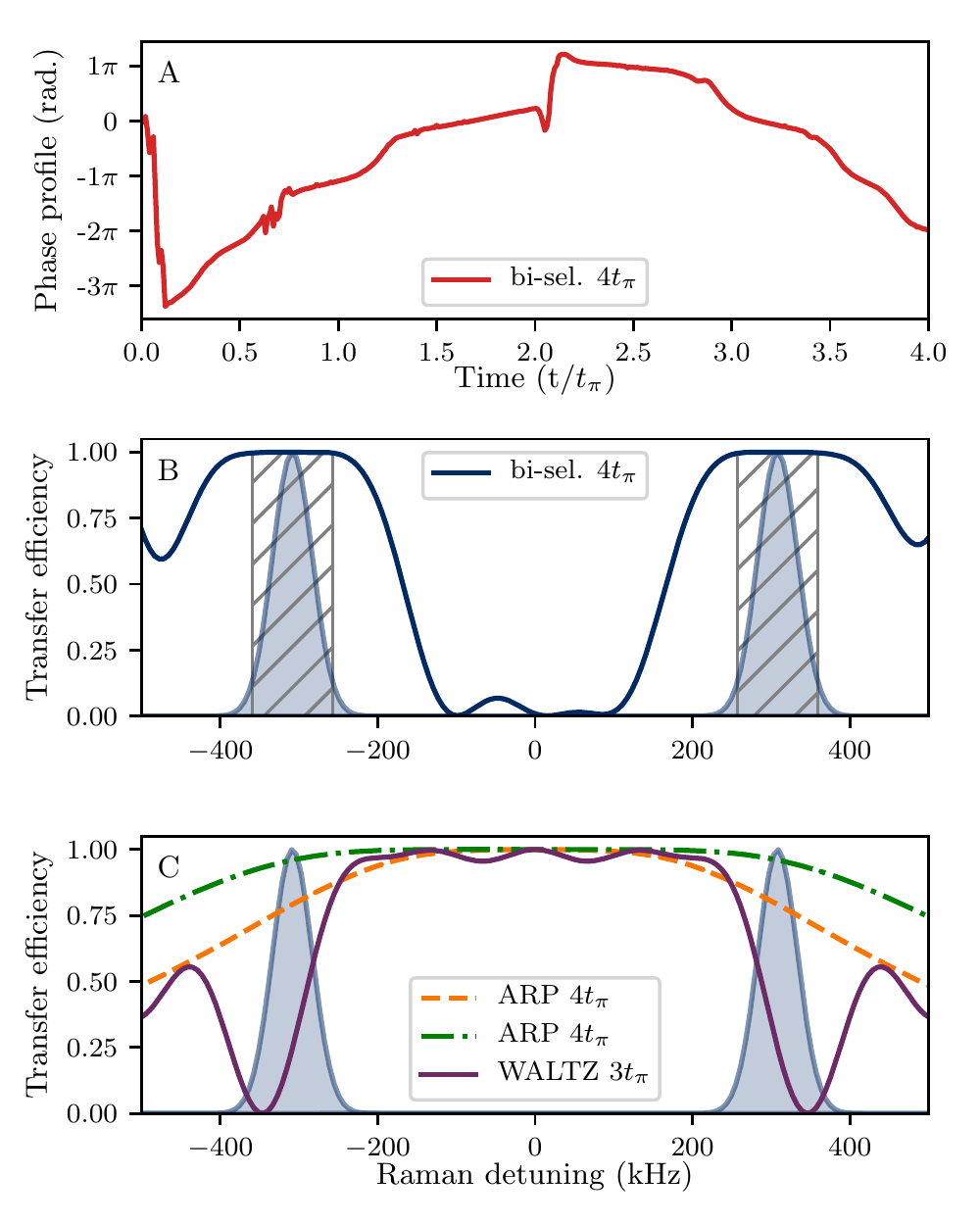}
  \caption{Phase profile (A) and simulated transfer efficiency (B) for the 10\textsuperscript{th} bi-selective pulse found optimising the fidelity $\mathcal{F}_{\mathrm{square}}$ (blue solid line). The transfer efficiency for the ARP Tanh/Tan pulse (orange dashed line and green dot-dashed line) and the WALTZ pulse (purple solid line) are shown in panel C. The detuning distributions of the two primary interferometer arms, corresponding to a temperature of 1 $\mu$K during the 10\textsuperscript{th} augmentation pulse are shown by shaded regions. The hatched regions represent the bands of detuning where the transfer efficiency is maximised using GRAPE.}
  \label{fig:bi-selective-spectral}
\end{figure}

Figure \ref{fig:bi-selective-spectral} shows the phase profile and simulated transfer efficiency of the 10\textsuperscript{th} augmentation pulse when the fidelity $\mathcal{F}_{\mathrm{square}}$ is instead optimised. Results are also shown for the WALTZ composite pulse and the Tanh/Tan ARP pulse with durations of 4$t_{\pi}$ and 8$t_{\pi}$. The corresponding detuning distributions for the upper and lower interferometer arms are also shown as shaded regions. Whereas the efficiencies of the WALTZ and ARP augmentation pulses are limited by the finite velocity acceptances, the bi-selective pulse is tailored for the split momentum distribution of the two primary interferometer arms, allowing it to achieve a higher efficiency than a Tanh/Tan ARP pulse of twice its duration. Many of the profiles found optimising $\mathcal{F}_{\mathrm{square}}$ have a strong parabolic component corresponding to a frequency chirp, although we are far from the adiabatic regime; and some include a $\pi$ phase step near the temporal midpoint: we find that this combination alone gives the main features but does not produce the flat regions of high efficiency achieved using optimal control. All the pulses found optimising $\mathcal{F}_{\mathrm{real}}$ are time-symmetric.

\subsection{LMT contrast}

We have simulated LMT interferometers of different LMT orders, and compared the performance using our bi-selective augmentation pulses with that using rectangular $\pi$ and Tanh/Tan ARP augmentation pulses. The contrast in each case is averaged over a thermal distribution of atoms and a uniform distribution of Rabi frequencies that is either kept temporally constant -- to represent a non-uniform laser intensity distribution -- or else varied
from pulse-to-pulse -- to represent the motion of atoms across such a
distribution. Following the approach taken by Kotru \cite{Kotru2016a},
only the primary interfering paths are included in the calculation to reduce
computation time.

\begin{figure*}[!htb]
  \centering \includegraphics[width=\linewidth]{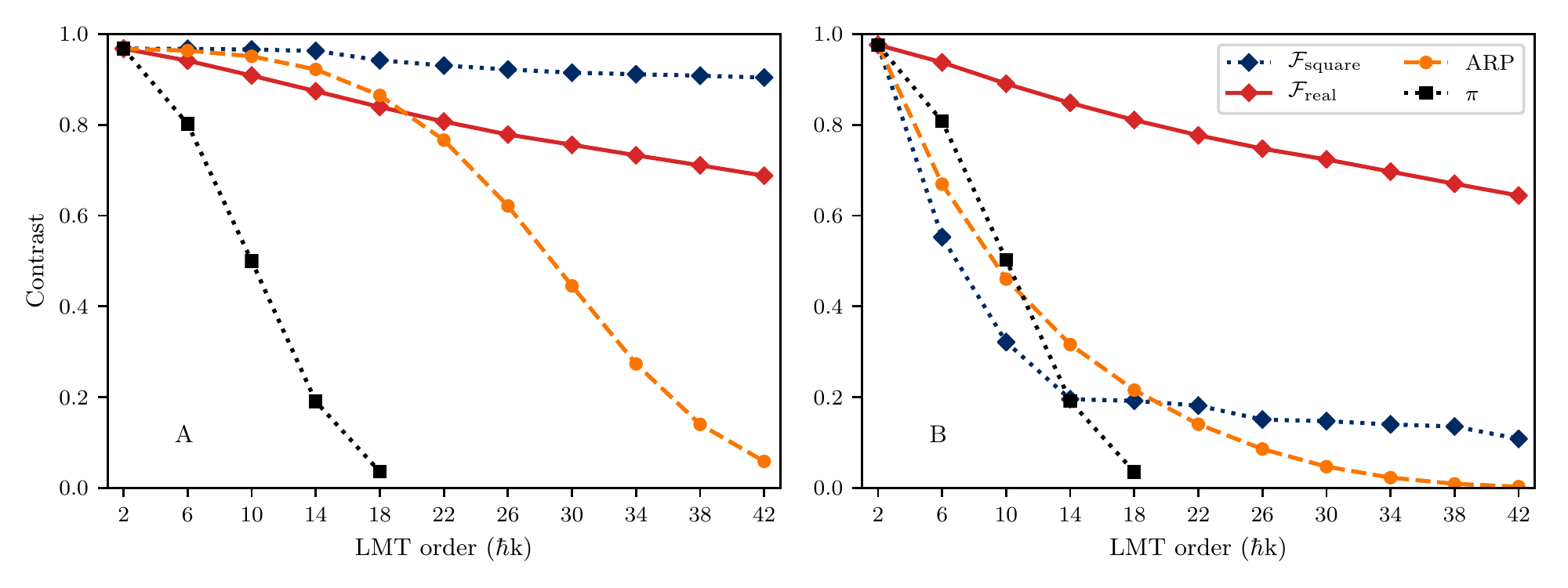}
  \caption{Simulated interferometer contrast for a cloud of $^{85}$Rb at 1$\mu$K as a function of LMT order for different pulse sequences. The contrast is numerically averaged over Raman
    detunings due to a velocity distribution corresponding to a
    temperature of 1$\mu$K and a uniform distribution of static Rabi rate errors
    of $\pm$10\% (A) or a random temporal variation of Rabi rate errors of
    $\pm$10\% from pulse-to-pulse (B). Contrast values for bi-selective GRAPE pulses
    found with fidelities $\mathcal{F}_{\mathrm{square}}$ and
    $\mathcal{F}_{\mathrm{real}}$ (blue dotted line with diamonds and red solid line with
    diamonds respectively) are shown. The results for $\pi$ augmentation pulses (black dotted line with rectangles) and the Tanh/Tan ARP pulse (dashed orange line with circles) are also shown.}
  \label{fig:bi-selective-contrast}
\end{figure*}

The simulation results  for an atomic temperature of 1 $\mu$K are shown in Figure \ref{fig:bi-selective-contrast}. To meaningfully compare the performance of the GRAPE bi-selective pulses with ARP, the total duration of
the sequences is kept the same. GRAPE bi-selective pulses far outperform basic rectangular $\pi$ pulses, and
Tanh/Tan ARP pulses of equivalent duration. When no thermal expansion is modelled, the dynamic phase
cancellation of the ARP pulses is perfect, meaning the reduction in contrast
with increasing LMT order is purely due to the limited velocity acceptance of
the pulses as the arms separate in detuning. For the ARP contrast to match that achievable with bi-selective pulses,
the ARP pulse duration must be increased, thus increasing the
susceptibility to spontaneous emission (not modelled currently) and dynamic phase dephasing \cite{Kotru2015}. This
highlights one key advantage of our adaptive approach.

If the Rabi frequency varies from pulse-to-pulse for different atoms, the dynamic phase imprinted on the atoms by ARP is not cancelled in the interferometer, meaning different atoms obtain
different phases and the interference is washed
out. Although ARP pulses have a velocity acceptance which increases with pulse duration, longer ARP pulses become more susceptible to errors in dynamic phase cancellation \cite{Kotru2015}.

When the Rabi rate is instead varied randomly between pulses in the simulation
to emulate noise in the Raman beam intensity profile (panel B in Figure
\ref{fig:bi-selective-contrast}), the bi-selective pulses found optimising the
phase-sensitive fidelity $\mathcal{F}_{\mathrm{real}}$ are able to maintain
significantly higher contrast than other pulse sequences. This is because the
phase variation of the wavepackets is minimised with respect to variations in
the Rabi frequency. This is not the case with pulses found using
$\mathcal{F}_{\mathrm{square}}$ and the contrast is significantly reduced in
such cases. 

We have repeated the bi-selective optimisation and contrast simulation for a hotter atomic temperature of 5 $\mu$K and for a longer pulse duration of $5t_{\pi}$. The results are summarised in Table \ref{table1}. We also show the results for the WALTZ composite pulse and a phase-modulated pulse obtained using GRAPE maximising $\mathcal{F}_{\mathrm{square}}$ for a large range of detunings centred on resonance. This non-selective pulse was optimised following the procedure outlined in \cite{Saywell2018b}. Longer pulses can achieve higher terminal fidelities, and using the phase-sensitive fidelity $\mathcal{F}_{\mathrm{real}}$ requires a longer duration than $\mathcal{F}_{\mathrm{square}}$ to reach an equivalent terminal fidelity. This is shown most clearly when optimising for the hotter cloud. There is no guarantee that we have found the global maximum for each choice of duration and temperature. Even so, pulses were found which outperform the composite and ARP alternatives tested.

\begin{table}[h]
\caption{ 
Simulated contrast values at temperatures of 1 and 5 $\mu$K for different LMT orders with no Rabi frequency variation between pulses. The $\pi$, WALTZ, and $\mathcal{F}_{\mathrm{square}}$ non-selective pulses are previous results with either defined ($\pi$, WALTZ) or previously chosen (non-selective) durations. The results for the ARP and bi-selective pulses are presented for two different durations. Bold values indicate the best performing pulse at each temperature.
} 
\footnotesize
\centering 
\begin{tabularx}{\linewidth}{l c ccc ccc}\hline\hline
  &\bf{Length} &  \multicolumn{3}{c}{\bf{1} $\mathbf{\mu}$\bf{K} \bf{Contrast}} & \multicolumn{3}{c}{\bf{5} $\mathbf{\mu}$\bf{K} \bf{Contrast}}  \\\cline{3-5}\cline{6-8}
 \raisebox{1.5ex}{\bf{Pulse}} & ($T/t_{\pi}$) &10$\hbar$k & 26$\hbar$k & 42$\hbar$k & 10$\hbar$k & 26$\hbar$k & 42$\hbar$k \\
\hline
$\pi$ & 1 & 0.50 & - & - & 0.35 & - & - \\[1.5ex]
WALTZ & 3 & 0.73 & 0.51 & 0.00 & 0.72 & 0.41 & 0.00 \\[1.5ex]

 $\mathcal{F}_{\mathrm{square}}$ non-selective&4&0.92&0.72&0.38&0.85&0.67&0.34\\[1.5ex]

 & 4& 0.95&0.62&0.06&0.88&0.52&0.04\\[-1ex]
\raisebox{1.5ex}{ARP} & 5&0.96&0.79&0.19&\bf{0.90}&0.69&0.15\\[1.5ex]

 & 4& 0.91&0.78&0.69&0.79&0.43&0.22\\[-1ex]
\raisebox{1.5ex}{$\mathcal{F}_{\mathrm{real}}$ bi-selective} & 5&0.95&0.87&0.82&0.85&0.60&0.50 \\[1.5ex]

 & 4& \bf{0.97}&0.92&0.90&0.83&0.64&0.57\\[-1ex]
\raisebox{1.5ex}{{$\mathcal{F}_{\mathrm{square}}$} bi-selective} & 5&0.96&\bf{0.94}&\bf{0.93}&0.89&\bf{0.81}&\bf{0.76} \\
\hline\hline
\end{tabularx}
\label{table1}
\end{table}

\subsection{Suppression of off-resonant excitation}
\label{sec:suppr-double-diffr}
\begin{figure*}[!htb]
  \centering \includegraphics[width=\linewidth]{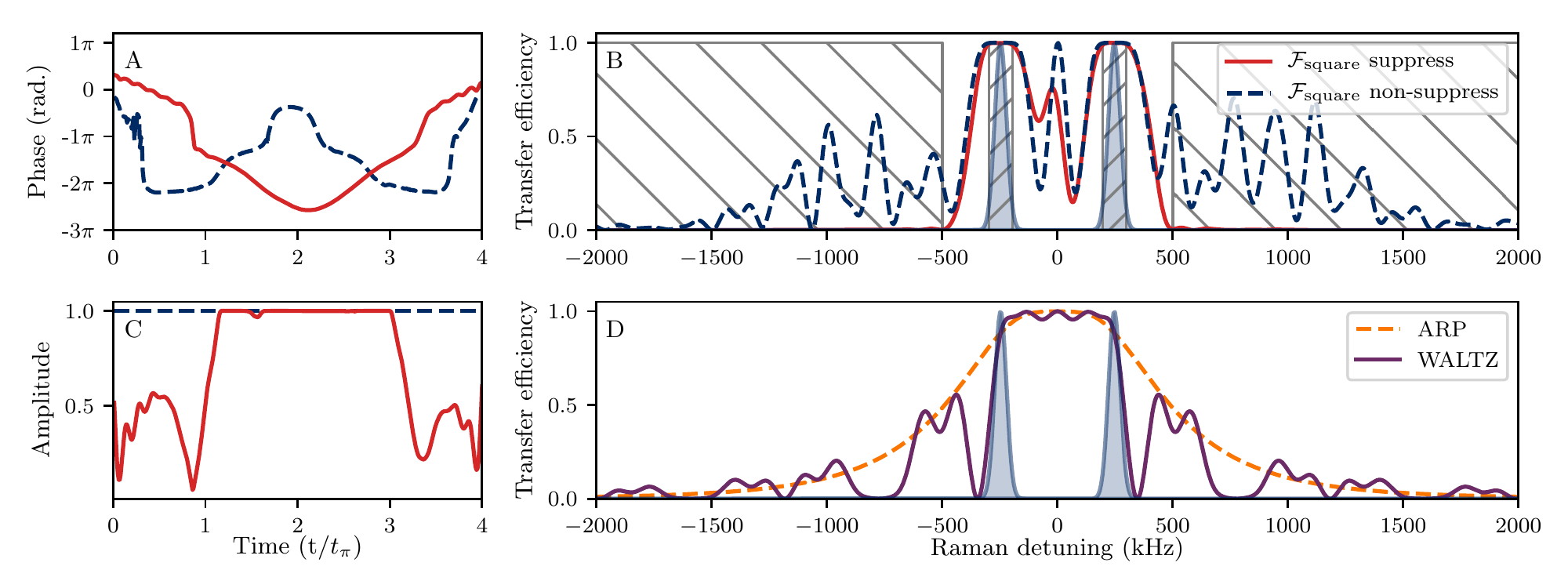}
  \caption{Panels A and C show the phase and amplitude profiles for two bi-selective pulses maximising $\mathcal{F}_{\mathrm{square}}$. The dashed blue curve shows the case where no suppression of off-resonant excitation is included in the optimisation and the solid red curve shows the case where off-resonant excitation is suppressed. The transfer efficiency of these two pulses as a function of detuning is shown in panel B, and panel D shows the transfer efficiency of the WALTZ and Tanh/Tan pulses. The diagonally hatched regions coinciding with the shaded detuning distributions show the optimisation bands of detuning during the 8\textsuperscript{th} augmentation pulse. The outer hatched regions show the range of the detunings for which the state transfer is suppressed ($\delta_{\mathrm{min}}, \delta_{\mathrm{max}} = 500,2500$ kHz). The response is unconstrained outside the hatched regions.}
  \label{fig:suppression-comparison}
\end{figure*}

When the Doppler shift from gravitational acceleration is used to discriminate between retroreflected frequency pairs in vertically-orientated interferometers, off-resonant excitations from broadband atom-optics can lead to unwanted double-diffraction. For example, following a 10 ms drop time \cite{Butts2013}, the resonance conditions of the two frequency pairs will shift by $2g\mathrm{k_{\mathrm{eff}}}\times 10\ \mathrm{ms} \approx$ 500 kHz for $^{85}$Rb, where $g$ is the local gravitational acceleration. Conventional robust pulses can result in a non-zero transfer at comparable detunings (Figure \ref{fig:suppression-comparison}), potentially
leading to double-diffraction if the unwanted frequency pair has not been shifted far enough from resonance \cite{Butts2013, Kotru2016a}.

We have optimised bi-selective pulses which suppress the transfer efficiency outside
the detuning bands of interest using a suitable modification of our bi-selective fidelity (Equation \ref{eq: bi-selective fidelity suppress}). Figure \ref{fig:suppression-comparison}
shows the preliminary results. In order to achieve a good level of suppression
at large detunings we have found it necessary to allow both the phase and
amplitude of the pulse to vary in the optimisation procedure. We have also added penalties in the optimisation to limit the maximum intensity during the pulse and enforce smooth waveforms \cite{Goodwin2015}. 

\section{Conclusions}
We have presented a new Raman pulse scheme for the augmentation pulses in large-area atom interferometry, whereby individually tailored Raman pulses found using optimal control techniques maintain resonance with the diverging wavepackets as the Raman detuning increases between the interferometer arms. We optimise our bi-selective pulses to provide maximum state-transfer while minimising variation in the interferometer phase across the atomic ensemble. The pulses can be made robust to large spatial and temporal variations in the Rabi rate,
potentially allowing LMT interferometry in non-ideal experimental environments such as those with warmer atom clouds and inhomogeneous laser beam fronts.
Our simulations show that our pulses can
maintain contrast at significantly higher momentum splittings than
interferometers that have employed the best augmentation
pulses demonstrated to-date, including the WALTZ composite pulse and the Tanh/Tan adiabatic rapid passage pulse, whose finite velocity acceptances limit the LMT momentum range. For large LMT orders, our
pulses can be considerably shorter than ARP or composite equivalents of the same
efficiency, reducing the susceptibility to spontaneous emission.

\appendix*
\section{Numerical model}
We model $^{85}$Rb atoms undergoing Raman transitions as two level systems,
described by the basis states $\ket{\mathrm{g}, \mathbf{p}}$ and
$\ket{\mathrm{e}, \mathbf{p} +\hbar\mathbf{k}_{\mathrm{eff}}}$ with
corresponding time-dependent amplitudes $c_{\mathrm{g,e}}(t)$. Pulses are
described by propagators of the form given by Equation \ref{eq: prop}, and are
given by time-ordered products of the propagators for individual slices in the
case of shaped and composite pulses, where the amplitude, detuning (frequency),
and phase may be varied for each step.

\begin{figure}[!htb]
  \centering \includegraphics[width=\linewidth]{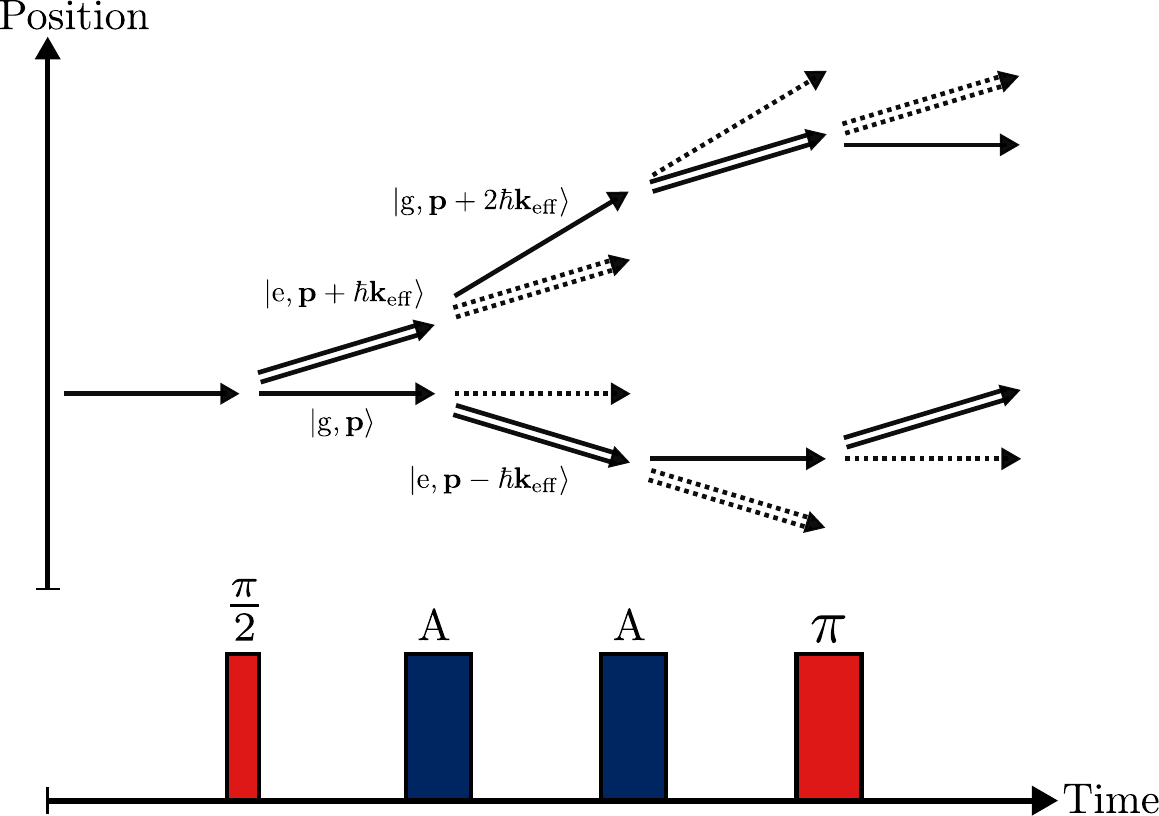}
  \caption{Diagram of the first half of an N=1 LMT pulse sequence, shown up to the central mirror pulse, indicating how the momentum and atomic state vary in each arm of the interferometer during the sequence. The resonant Raman frequency depends upon the pair of states coupled at each stage. `A' represents an augmentation pulse. Single and double arrows represent the internal states $\ket{\mathrm{g}}$ and $\ket{\mathrm{e}}$ respectively.}
  \label{fig:LMT_diagram2}
\end{figure}

The basis states for each pulse in an LMT sequence vary. For example, during the
initial $\pi/2$ pulse the basis is $\ket{\mathrm{g},
  \mathbf{p}}$,$\ket{\mathrm{e}, \mathbf{p} +\hbar\mathbf{k}_{\mathrm{eff}}}$,
but for the first augmentation pulse (where the wavevector is reversed) the
upper arm has a basis described by the states $\ket{\mathrm{g},
  \mathbf{p}+2\hbar\mathbf{k}_{\mathrm{eff}}}$,$\ket{\mathrm{e}, \mathbf{p}
  +\hbar\mathbf{k}_{\mathrm{eff}}}$, and the lower arm has a basis described by
the states $\ket{\mathrm{g}, \mathbf{p}}$,$\ket{\mathrm{e}, \mathbf{p}
  -\hbar\mathbf{k}_{\mathrm{eff}}}$. This means the resonance conditions are
separated for the two interferometer arms by 4$\delta_{\mathrm{recoil}}$.
Furthermore, to reduce computation time, only the amplitudes following the two primary interferometer arms
are included in the calculation, as shown in Figure
\ref{fig:LMT_diagram2}. This is a good assumption when the pulses providing the
large-momentum transfer are efficient, or when there is no interference from atoms that have not followed the primary interferometer paths \cite{Kotru2016a}.

We simulate how the interferometer contrast is affected by the atomic
temperature and variations in the Rabi rate across the atom cloud. In order to
do this, we draw a sample of atomic velocities (and hence Raman detunings) from
a Maxwell-Boltzmann distribution for $^{85}$Rb at a given temperature and a
uniform distribution of Rabi frequencies in the range $\pm$10\% of the
intended Rabi rate. Alternatively, the Rabi rate may be randomly varied from pulse-to-pulse within $\pm$10\% of the
intended Rabi rate in order to explore the robustness of sequences to temporal variation throughout the interferometer.

The contrast following an interferometer pulse sequence may be calculated by
evolving the state amplitudes for an atom initially in the ground internal
state, $\ket{\mathrm{g}}$, by applying the relevant pulse propagators in the
correct order with the correct Raman detunings for each pulse and interferometer
arm. Fringes were calculated for each atom in the ensemble by repeating the evolution of the final rectangular beamsplitter and varying the phase, $\phi_{bs}$, between 0 and $2\pi$. The ensemble average interferograms were fitted to the sinusoidal function $0.5(A+B\cos(\phi_{bs} +
C))$, where $A$ is an offset, $B$ is the contrast, and $C$ is a possible phase
shift. The effect of spontaneous emission
is not modelled at present but will ultimately limit the achievable contrast
with extended pulse sequences in LMT interferometers.

The sweep parameters used to define the Tanh/Tan ARP pulse are the same as those used by Kotru \textit{et al.} \cite{Kotru2015}. The Tanh/Tan ARP pulse waveforms were divided into 2500 time-steps in the simulations to ensure the piece-wise constant approximation was sufficiently accurate to model the frequency and amplitude sweep.

\begin{acknowledgments}
  This work was supported by Dstl (Grant Nos. DSTLX-1000091758
  and DSTLX-1000097855) and the UK Engineering and Physical Sciences Research
  Council (Grant Nos. EP/M013294/1 and EP/L015382/1).
\end{acknowledgments}
	
\bibliography{selective_pulses}

\begin{thebibliography}{54}%
\makeatletter
\providecommand \@ifxundefined [1]{%
 \@ifx{#1\undefined}
}%
\providecommand \@ifnum [1]{%
 \ifnum #1\expandafter \@firstoftwo
 \else \expandafter \@secondoftwo
 \fi
}%
\providecommand \@ifx [1]{%
 \ifx #1\expandafter \@firstoftwo
 \else \expandafter \@secondoftwo
 \fi
}%
\providecommand \natexlab [1]{#1}%
\providecommand \enquote  [1]{``#1''}%
\providecommand \bibnamefont  [1]{#1}%
\providecommand \bibfnamefont [1]{#1}%
\providecommand \citenamefont [1]{#1}%
\providecommand \href@noop [0]{\@secondoftwo}%
\providecommand \href [0]{\begingroup \@sanitize@url \@href}%
\providecommand \@href[1]{\@@startlink{#1}\@@href}%
\providecommand \@@href[1]{\endgroup#1\@@endlink}%
\providecommand \@sanitize@url [0]{\catcode `\\12\catcode `\$12\catcode
  `\&12\catcode `\#12\catcode `\^12\catcode `\_12\catcode `\%12\relax}%
\providecommand \@@startlink[1]{}%
\providecommand \@@endlink[0]{}%
\providecommand \url  [0]{\begingroup\@sanitize@url \@url }%
\providecommand \@url [1]{\endgroup\@href {#1}{\urlprefix }}%
\providecommand \urlprefix  [0]{URL }%
\providecommand \Eprint [0]{\href }%
\providecommand \doibase [0]{http://dx.doi.org/}%
\providecommand \selectlanguage [0]{\@gobble}%
\providecommand \bibinfo  [0]{\@secondoftwo}%
\providecommand \bibfield  [0]{\@secondoftwo}%
\providecommand \translation [1]{[#1]}%
\providecommand \BibitemOpen [0]{}%
\providecommand \bibitemStop [0]{}%
\providecommand \bibitemNoStop [0]{.\EOS\space}%
\providecommand \EOS [0]{\spacefactor3000\relax}%
\providecommand \BibitemShut  [1]{\csname bibitem#1\endcsname}%
\let\auto@bib@innerbib\@empty
\bibitem [{\citenamefont {Berman}(1997)}]{RBerman1997}%
  \BibitemOpen
  \bibinfo {editor} {\bibfnamefont {P.}~\bibnamefont {Berman}},\ ed.,\ \href
  {https://www.sciencedirect.com/book/9780120924608/atom-interferometry} {\emph
  {\bibinfo {title} {{Atom Interferometry}}}}\ (\bibinfo  {publisher} {Academic
  Press},\ \bibinfo {year} {1997})\ p.\ \bibinfo {pages} {363}\BibitemShut
  {NoStop}%
\bibitem [{\citenamefont {Peters}\ \emph {et~al.}(2001)\citenamefont {Peters},
  \citenamefont {Chung},\ and\ \citenamefont {Chu}}]{Peters2001}%
  \BibitemOpen
  \bibfield  {author} {\bibinfo {author} {\bibfnamefont {A.}~\bibnamefont
  {Peters}}, \bibinfo {author} {\bibfnamefont {K.~Y.}\ \bibnamefont {Chung}}, \
  and\ \bibinfo {author} {\bibfnamefont {S.}~\bibnamefont {Chu}},\ }\href
  {\doibase 10.1088/0026-1394/38/1/4} {\bibfield  {journal} {\bibinfo
  {journal} {Metrologia}\ }\textbf {\bibinfo {volume} {38}},\ \bibinfo {pages}
  {25} (\bibinfo {year} {2001})}\BibitemShut {NoStop}%
\bibitem [{\citenamefont {{Mc Guirk}}\ \emph {et~al.}(2002)\citenamefont {{Mc
  Guirk}}, \citenamefont {Foster}, \citenamefont {Fixler}, \citenamefont
  {Snadden},\ and\ \citenamefont {Kasevich}}]{Mcguirk2002}%
  \BibitemOpen
  \bibfield  {author} {\bibinfo {author} {\bibfnamefont {J.~M.}\ \bibnamefont
  {{Mc Guirk}}}, \bibinfo {author} {\bibfnamefont {G.~T.}\ \bibnamefont
  {Foster}}, \bibinfo {author} {\bibfnamefont {J.~B.}\ \bibnamefont {Fixler}},
  \bibinfo {author} {\bibfnamefont {M.~J.}\ \bibnamefont {Snadden}}, \ and\
  \bibinfo {author} {\bibfnamefont {M.~A.}\ \bibnamefont {Kasevich}},\ }\href
  {\doibase 10.1103/PhysRevA.65.033608} {\bibfield  {journal} {\bibinfo
  {journal} {Phys. Rev. A}\ }\textbf {\bibinfo {volume} {65}},\ \bibinfo
  {pages} {033608} (\bibinfo {year} {2002})}\BibitemShut {NoStop}%
\bibitem [{\citenamefont {Rosi}\ \emph {et~al.}(2014)\citenamefont {Rosi},
  \citenamefont {Sorrentino}, \citenamefont {Cacciapuoti}, \citenamefont
  {Prevedelli},\ and\ \citenamefont {Tino}}]{Rosi2014}%
  \BibitemOpen
  \bibfield  {author} {\bibinfo {author} {\bibfnamefont {G.}~\bibnamefont
  {Rosi}}, \bibinfo {author} {\bibfnamefont {F.}~\bibnamefont {Sorrentino}},
  \bibinfo {author} {\bibfnamefont {L.}~\bibnamefont {Cacciapuoti}}, \bibinfo
  {author} {\bibfnamefont {M.}~\bibnamefont {Prevedelli}}, \ and\ \bibinfo
  {author} {\bibfnamefont {G.~M.}\ \bibnamefont {Tino}},\ }\href {\doibase
  10.1038/nature13433} {\bibfield  {journal} {\bibinfo  {journal} {Nature}\
  }\textbf {\bibinfo {volume} {510}},\ \bibinfo {pages} {518} (\bibinfo {year}
  {2014})}\BibitemShut {NoStop}%
\bibitem [{\citenamefont {Gustavson}\ \emph {et~al.}(1997)\citenamefont
  {Gustavson}, \citenamefont {Bouyer},\ and\ \citenamefont
  {Kasevich}}]{Gustavson1997}%
  \BibitemOpen
  \bibfield  {author} {\bibinfo {author} {\bibfnamefont {T.~L.}\ \bibnamefont
  {Gustavson}}, \bibinfo {author} {\bibfnamefont {P.}~\bibnamefont {Bouyer}}, \
  and\ \bibinfo {author} {\bibfnamefont {M.~A.}\ \bibnamefont {Kasevich}},\
  }\href {\doibase 10.1103/PhysRevLett.78.2046} {\bibfield  {journal} {\bibinfo
   {journal} {Phys. Rev. Lett.}\ }\textbf {\bibinfo {volume} {78}},\ \bibinfo
  {pages} {2046} (\bibinfo {year} {1997})}\BibitemShut {NoStop}%
\bibitem [{\citenamefont {Barrett}\ \emph {et~al.}(2014)\citenamefont
  {Barrett}, \citenamefont {Geiger}, \citenamefont {Dutta}, \citenamefont
  {Meunier}, \citenamefont {Canuel}, \citenamefont {Gauguet}, \citenamefont
  {Bouyer},\ and\ \citenamefont {Landragin}}]{Barrett2014}%
  \BibitemOpen
  \bibfield  {author} {\bibinfo {author} {\bibfnamefont {B.}~\bibnamefont
  {Barrett}}, \bibinfo {author} {\bibfnamefont {R.}~\bibnamefont {Geiger}},
  \bibinfo {author} {\bibfnamefont {I.}~\bibnamefont {Dutta}}, \bibinfo
  {author} {\bibfnamefont {M.}~\bibnamefont {Meunier}}, \bibinfo {author}
  {\bibfnamefont {B.}~\bibnamefont {Canuel}}, \bibinfo {author} {\bibfnamefont
  {A.}~\bibnamefont {Gauguet}}, \bibinfo {author} {\bibfnamefont
  {P.}~\bibnamefont {Bouyer}}, \ and\ \bibinfo {author} {\bibfnamefont
  {A.}~\bibnamefont {Landragin}},\ }\href {\doibase 10.1016/j.crhy.2014.10.009}
  {\bibfield  {journal} {\bibinfo  {journal} {C. R. Phys.}\ }\textbf {\bibinfo
  {volume} {15}},\ \bibinfo {pages} {875} (\bibinfo {year} {2014})}\BibitemShut
  {NoStop}%
\bibitem [{\citenamefont {Hoth}\ \emph {et~al.}(2016)\citenamefont {Hoth},
  \citenamefont {Pelle}, \citenamefont {Riedl}, \citenamefont {Kitching},\ and\
  \citenamefont {Donley}}]{Hoth2016}%
  \BibitemOpen
  \bibfield  {author} {\bibinfo {author} {\bibfnamefont {G.~W.}\ \bibnamefont
  {Hoth}}, \bibinfo {author} {\bibfnamefont {B.}~\bibnamefont {Pelle}},
  \bibinfo {author} {\bibfnamefont {S.}~\bibnamefont {Riedl}}, \bibinfo
  {author} {\bibfnamefont {J.}~\bibnamefont {Kitching}}, \ and\ \bibinfo
  {author} {\bibfnamefont {E.~A.}\ \bibnamefont {Donley}},\ }\href {\doibase
  10.1063/1.4961527} {\bibfield  {journal} {\bibinfo  {journal} {Appl. Phys.
  Lett.}\ }\textbf {\bibinfo {volume} {109}},\ \bibinfo {pages} {071113}
  (\bibinfo {year} {2016})}\BibitemShut {NoStop}%
\bibitem [{\citenamefont {Hamilton}\ \emph {et~al.}(2015)\citenamefont
  {Hamilton}, \citenamefont {Jaffe}, \citenamefont {Haslinger}, \citenamefont
  {Simmons}, \citenamefont {Muller},\ and\ \citenamefont
  {Khoury}}]{Hamilton2015a}%
  \BibitemOpen
  \bibfield  {author} {\bibinfo {author} {\bibfnamefont {P.}~\bibnamefont
  {Hamilton}}, \bibinfo {author} {\bibfnamefont {M.}~\bibnamefont {Jaffe}},
  \bibinfo {author} {\bibfnamefont {P.}~\bibnamefont {Haslinger}}, \bibinfo
  {author} {\bibfnamefont {Q.}~\bibnamefont {Simmons}}, \bibinfo {author}
  {\bibfnamefont {H.}~\bibnamefont {Muller}}, \ and\ \bibinfo {author}
  {\bibfnamefont {J.}~\bibnamefont {Khoury}},\ }\href {\doibase
  10.1126/science.aaa8883} {\bibfield  {journal} {\bibinfo  {journal}
  {Science}\ }\textbf {\bibinfo {volume} {349}},\ \bibinfo {pages} {849}
  (\bibinfo {year} {2015})}\BibitemShut {NoStop}%
\bibitem [{\citenamefont {Jaffe}\ \emph {et~al.}(2017)\citenamefont {Jaffe},
  \citenamefont {Haslinger}, \citenamefont {Xu}, \citenamefont {Hamilton},
  \citenamefont {Upadhye}, \citenamefont {Elder}, \citenamefont {Khoury},\ and\
  \citenamefont {M{\"{u}}ller}}]{Jaffe2017}%
  \BibitemOpen
  \bibfield  {author} {\bibinfo {author} {\bibfnamefont {M.}~\bibnamefont
  {Jaffe}}, \bibinfo {author} {\bibfnamefont {P.}~\bibnamefont {Haslinger}},
  \bibinfo {author} {\bibfnamefont {V.}~\bibnamefont {Xu}}, \bibinfo {author}
  {\bibfnamefont {P.}~\bibnamefont {Hamilton}}, \bibinfo {author}
  {\bibfnamefont {A.}~\bibnamefont {Upadhye}}, \bibinfo {author} {\bibfnamefont
  {B.}~\bibnamefont {Elder}}, \bibinfo {author} {\bibfnamefont
  {J.}~\bibnamefont {Khoury}}, \ and\ \bibinfo {author} {\bibfnamefont
  {H.}~\bibnamefont {M{\"{u}}ller}},\ }\href {\doibase 10.1038/nphys4189}
  {\bibfield  {journal} {\bibinfo  {journal} {Nat. Phys.}\ }\textbf {\bibinfo
  {volume} {13}},\ \bibinfo {pages} {938} (\bibinfo {year} {2017})}\BibitemShut
  {NoStop}%
\bibitem [{\citenamefont {M{\"{u}}ller}\ \emph {et~al.}(2008)\citenamefont
  {M{\"{u}}ller}, \citenamefont {Chiow}, \citenamefont {Long}, \citenamefont
  {Herrmann},\ and\ \citenamefont {Chu}}]{Muller2008c}%
  \BibitemOpen
  \bibfield  {author} {\bibinfo {author} {\bibfnamefont {H.}~\bibnamefont
  {M{\"{u}}ller}}, \bibinfo {author} {\bibfnamefont {S.-w.}\ \bibnamefont
  {Chiow}}, \bibinfo {author} {\bibfnamefont {Q.}~\bibnamefont {Long}},
  \bibinfo {author} {\bibfnamefont {S.}~\bibnamefont {Herrmann}}, \ and\
  \bibinfo {author} {\bibfnamefont {S.}~\bibnamefont {Chu}},\ }\href {\doibase
  10.1103/PhysRevLett.100.180405} {\bibfield  {journal} {\bibinfo  {journal}
  {Phys. Rev. Lett.}\ }\textbf {\bibinfo {volume} {100}},\ \bibinfo {pages}
  {180405} (\bibinfo {year} {2008})}\BibitemShut {NoStop}%
\bibitem [{\citenamefont {Kasevich}\ and\ \citenamefont
  {Chu}(1991)}]{Kasevich1991a}%
  \BibitemOpen
  \bibfield  {author} {\bibinfo {author} {\bibfnamefont {M.}~\bibnamefont
  {Kasevich}}\ and\ \bibinfo {author} {\bibfnamefont {S.}~\bibnamefont {Chu}},\
  }\href {\doibase 10.1103/PhysRevLett.67.181} {\bibfield  {journal} {\bibinfo
  {journal} {Phys. Rev. Lett.}\ }\textbf {\bibinfo {volume} {67}},\ \bibinfo
  {pages} {181} (\bibinfo {year} {1991})}\BibitemShut {NoStop}%
\bibitem [{\citenamefont {Kasevich}\ \emph {et~al.}(1991)\citenamefont
  {Kasevich}, \citenamefont {Weiss}, \citenamefont {Riis}, \citenamefont
  {Moler}, \citenamefont {Kasapi},\ and\ \citenamefont {Chu}}]{Kasevich1991b}%
  \BibitemOpen
  \bibfield  {author} {\bibinfo {author} {\bibfnamefont {M.}~\bibnamefont
  {Kasevich}}, \bibinfo {author} {\bibfnamefont {D.~S.}\ \bibnamefont {Weiss}},
  \bibinfo {author} {\bibfnamefont {E.}~\bibnamefont {Riis}}, \bibinfo {author}
  {\bibfnamefont {K.}~\bibnamefont {Moler}}, \bibinfo {author} {\bibfnamefont
  {S.}~\bibnamefont {Kasapi}}, \ and\ \bibinfo {author} {\bibfnamefont
  {S.}~\bibnamefont {Chu}},\ }\href {\doibase 10.1103/PhysRevLett.66.2297}
  {\bibfield  {journal} {\bibinfo  {journal} {Phys. Rev. Lett.}\ }\textbf
  {\bibinfo {volume} {66}},\ \bibinfo {pages} {2297} (\bibinfo {year}
  {1991})}\BibitemShut {NoStop}%
\bibitem [{\citenamefont {McGuirk}\ \emph {et~al.}(2000)\citenamefont
  {McGuirk}, \citenamefont {Snadden},\ and\ \citenamefont
  {Kasevich}}]{McGuirk2000}%
  \BibitemOpen
  \bibfield  {author} {\bibinfo {author} {\bibfnamefont {J.~M.}\ \bibnamefont
  {McGuirk}}, \bibinfo {author} {\bibfnamefont {M.~J.}\ \bibnamefont
  {Snadden}}, \ and\ \bibinfo {author} {\bibfnamefont {M.~A.}\ \bibnamefont
  {Kasevich}},\ }\href {\doibase 10.1103/PhysRevLett.85.4498} {\bibfield
  {journal} {\bibinfo  {journal} {Phys. Rev. Lett.}\ }\textbf {\bibinfo
  {volume} {85}},\ \bibinfo {pages} {4498} (\bibinfo {year}
  {2000})}\BibitemShut {NoStop}%
\bibitem [{\citenamefont {Butts}\ \emph {et~al.}(2013)\citenamefont {Butts},
  \citenamefont {Kotru}, \citenamefont {Kinast}, \citenamefont {Radojevic},
  \citenamefont {Timmons},\ and\ \citenamefont {Stoner}}]{Butts2013}%
  \BibitemOpen
  \bibfield  {author} {\bibinfo {author} {\bibfnamefont {D.~L.}\ \bibnamefont
  {Butts}}, \bibinfo {author} {\bibfnamefont {K.}~\bibnamefont {Kotru}},
  \bibinfo {author} {\bibfnamefont {J.~M.}\ \bibnamefont {Kinast}}, \bibinfo
  {author} {\bibfnamefont {A.~M.}\ \bibnamefont {Radojevic}}, \bibinfo {author}
  {\bibfnamefont {B.~P.}\ \bibnamefont {Timmons}}, \ and\ \bibinfo {author}
  {\bibfnamefont {R.~E.}\ \bibnamefont {Stoner}},\ }\href {\doibase
  10.1364/JOSAB.30.000922} {\bibfield  {journal} {\bibinfo  {journal} {J. Opt.
  Soc. Am. B}\ }\textbf {\bibinfo {volume} {30}},\ \bibinfo {pages} {922}
  (\bibinfo {year} {2013})}\BibitemShut {NoStop}%
\bibitem [{\citenamefont {Kotru}\ \emph {et~al.}(2015)\citenamefont {Kotru},
  \citenamefont {Butts}, \citenamefont {Kinast},\ and\ \citenamefont
  {Stoner}}]{Kotru2015}%
  \BibitemOpen
  \bibfield  {author} {\bibinfo {author} {\bibfnamefont {K.}~\bibnamefont
  {Kotru}}, \bibinfo {author} {\bibfnamefont {D.~L.}\ \bibnamefont {Butts}},
  \bibinfo {author} {\bibfnamefont {J.~M.}\ \bibnamefont {Kinast}}, \ and\
  \bibinfo {author} {\bibfnamefont {R.~E.}\ \bibnamefont {Stoner}},\ }\href
  {\doibase 10.1103/PhysRevLett.115.103001} {\bibfield  {journal} {\bibinfo
  {journal} {Phys. Rev. Lett.}\ }\textbf {\bibinfo {volume} {115}},\ \bibinfo
  {pages} {103001} (\bibinfo {year} {2015})}\BibitemShut {NoStop}%
\bibitem [{\citenamefont {Rudolph}\ \emph {et~al.}(2020)\citenamefont
  {Rudolph}, \citenamefont {Wilkason}, \citenamefont {Nantel}, \citenamefont
  {Swan}, \citenamefont {Holland}, \citenamefont {Jiang}, \citenamefont
  {Garber}, \citenamefont {Carman},\ and\ \citenamefont {Hogan}}]{Rudolph2020}%
  \BibitemOpen
  \bibfield  {author} {\bibinfo {author} {\bibfnamefont {J.}~\bibnamefont
  {Rudolph}}, \bibinfo {author} {\bibfnamefont {T.}~\bibnamefont {Wilkason}},
  \bibinfo {author} {\bibfnamefont {M.}~\bibnamefont {Nantel}}, \bibinfo
  {author} {\bibfnamefont {H.}~\bibnamefont {Swan}}, \bibinfo {author}
  {\bibfnamefont {C.~M.}\ \bibnamefont {Holland}}, \bibinfo {author}
  {\bibfnamefont {Y.}~\bibnamefont {Jiang}}, \bibinfo {author} {\bibfnamefont
  {B.~E.}\ \bibnamefont {Garber}}, \bibinfo {author} {\bibfnamefont {S.~P.}\
  \bibnamefont {Carman}}, \ and\ \bibinfo {author} {\bibfnamefont {J.~M.}\
  \bibnamefont {Hogan}},\ }\href {\doibase 10.1103/PhysRevLett.124.083604}
  {\bibfield  {journal} {\bibinfo  {journal} {Phys. Rev. Lett.}\ }\textbf
  {\bibinfo {volume} {124}},\ \bibinfo {pages} {083604} (\bibinfo {year}
  {2020})}\BibitemShut {NoStop}%
\bibitem [{\citenamefont {Mielec}\ \emph {et~al.}(2018)\citenamefont {Mielec},
  \citenamefont {Altorio}, \citenamefont {Sapam}, \citenamefont {Horville},
  \citenamefont {Holleville}, \citenamefont {Sidorenkov}, \citenamefont
  {Landragin},\ and\ \citenamefont {Geiger}}]{Mielec2018}%
  \BibitemOpen
  \bibfield  {author} {\bibinfo {author} {\bibfnamefont {N.}~\bibnamefont
  {Mielec}}, \bibinfo {author} {\bibfnamefont {M.}~\bibnamefont {Altorio}},
  \bibinfo {author} {\bibfnamefont {R.}~\bibnamefont {Sapam}}, \bibinfo
  {author} {\bibfnamefont {D.}~\bibnamefont {Horville}}, \bibinfo {author}
  {\bibfnamefont {D.}~\bibnamefont {Holleville}}, \bibinfo {author}
  {\bibfnamefont {L.~A.}\ \bibnamefont {Sidorenkov}}, \bibinfo {author}
  {\bibfnamefont {A.}~\bibnamefont {Landragin}}, \ and\ \bibinfo {author}
  {\bibfnamefont {R.}~\bibnamefont {Geiger}},\ }\href {\doibase
  10.1063/1.5051663} {\bibfield  {journal} {\bibinfo  {journal} {Appl. Phys.
  Lett.}\ }\textbf {\bibinfo {volume} {113}},\ \bibinfo {pages} {161108}
  (\bibinfo {year} {2018})}\BibitemShut {NoStop}%
\bibitem [{\citenamefont {Levitt}\ and\ \citenamefont
  {Freeman}(1981)}]{Levitt1981a}%
  \BibitemOpen
  \bibfield  {author} {\bibinfo {author} {\bibfnamefont {M.~H.}\ \bibnamefont
  {Levitt}}\ and\ \bibinfo {author} {\bibfnamefont {R.}~\bibnamefont
  {Freeman}},\ }\href {\doibase 10.1016/0022-2364(81)90082-2} {\bibfield
  {journal} {\bibinfo  {journal} {J. Magn. Reson. (1969)}\ }\textbf {\bibinfo
  {volume} {43}},\ \bibinfo {pages} {65} (\bibinfo {year} {1981})}\BibitemShut
  {NoStop}%
\bibitem [{\citenamefont {Dunning}\ \emph {et~al.}(2014)\citenamefont
  {Dunning}, \citenamefont {Gregory}, \citenamefont {Bateman}, \citenamefont
  {Cooper}, \citenamefont {Himsworth}, \citenamefont {Jones},\ and\
  \citenamefont {Freegarde}}]{Dunning2014}%
  \BibitemOpen
  \bibfield  {author} {\bibinfo {author} {\bibfnamefont {A.}~\bibnamefont
  {Dunning}}, \bibinfo {author} {\bibfnamefont {R.}~\bibnamefont {Gregory}},
  \bibinfo {author} {\bibfnamefont {J.}~\bibnamefont {Bateman}}, \bibinfo
  {author} {\bibfnamefont {N.}~\bibnamefont {Cooper}}, \bibinfo {author}
  {\bibfnamefont {M.}~\bibnamefont {Himsworth}}, \bibinfo {author}
  {\bibfnamefont {J.~A.}\ \bibnamefont {Jones}}, \ and\ \bibinfo {author}
  {\bibfnamefont {T.}~\bibnamefont {Freegarde}},\ }\href {\doibase
  10.1103/PhysRevA.90.033608} {\bibfield  {journal} {\bibinfo  {journal} {Phys.
  Rev. A}\ }\textbf {\bibinfo {volume} {90}},\ \bibinfo {pages} {033608}
  (\bibinfo {year} {2014})}\BibitemShut {NoStop}%
\bibitem [{\citenamefont {Bateman}\ and\ \citenamefont
  {Freegarde}(2007)}]{Bateman2007}%
  \BibitemOpen
  \bibfield  {author} {\bibinfo {author} {\bibfnamefont {J.}~\bibnamefont
  {Bateman}}\ and\ \bibinfo {author} {\bibfnamefont {T.}~\bibnamefont
  {Freegarde}},\ }\href {\doibase 10.1103/PhysRevA.76.013416} {\bibfield
  {journal} {\bibinfo  {journal} {Phys. Rev. A}\ }\textbf {\bibinfo {volume}
  {76}},\ \bibinfo {pages} {013416} (\bibinfo {year} {2007})}\BibitemShut
  {NoStop}%
\bibitem [{\citenamefont {Chiow}\ \emph {et~al.}(2011)\citenamefont {Chiow},
  \citenamefont {Kovachy}, \citenamefont {Chien},\ and\ \citenamefont
  {Kasevich}}]{Chiow2011}%
  \BibitemOpen
  \bibfield  {author} {\bibinfo {author} {\bibfnamefont {S.~W.}\ \bibnamefont
  {Chiow}}, \bibinfo {author} {\bibfnamefont {T.}~\bibnamefont {Kovachy}},
  \bibinfo {author} {\bibfnamefont {H.~C.}\ \bibnamefont {Chien}}, \ and\
  \bibinfo {author} {\bibfnamefont {M.~A.}\ \bibnamefont {Kasevich}},\ }\href
  {\doibase 10.1103/PhysRevLett.107.130403} {\bibfield  {journal} {\bibinfo
  {journal} {Phys. Rev. Lett.}\ }\textbf {\bibinfo {volume} {107}},\ \bibinfo
  {pages} {130403} (\bibinfo {year} {2011})}\BibitemShut {NoStop}%
\bibitem [{\citenamefont {Clad{\'{e}}}\ \emph {et~al.}(2009)\citenamefont
  {Clad{\'{e}}}, \citenamefont {Guellati-Kh{\'{e}}lifa}, \citenamefont {Nez},\
  and\ \citenamefont {Biraben}}]{Clade2009}%
  \BibitemOpen
  \bibfield  {author} {\bibinfo {author} {\bibfnamefont {P.}~\bibnamefont
  {Clad{\'{e}}}}, \bibinfo {author} {\bibfnamefont {S.}~\bibnamefont
  {Guellati-Kh{\'{e}}lifa}}, \bibinfo {author} {\bibfnamefont {F.}~\bibnamefont
  {Nez}}, \ and\ \bibinfo {author} {\bibfnamefont {F.}~\bibnamefont
  {Biraben}},\ }\href {\doibase 10.1103/PhysRevLett.102.240402} {\bibfield
  {journal} {\bibinfo  {journal} {Phys. Rev. Lett.}\ }\textbf {\bibinfo
  {volume} {102}},\ \bibinfo {pages} {240402} (\bibinfo {year}
  {2009})}\BibitemShut {NoStop}%
\bibitem [{\citenamefont {M{\"{u}}ller}\ \emph {et~al.}(2009)\citenamefont
  {M{\"{u}}ller}, \citenamefont {Chiow}, \citenamefont {Herrmann},\ and\
  \citenamefont {Chu}}]{Muller2009}%
  \BibitemOpen
  \bibfield  {author} {\bibinfo {author} {\bibfnamefont {H.}~\bibnamefont
  {M{\"{u}}ller}}, \bibinfo {author} {\bibfnamefont {S.-w.}\ \bibnamefont
  {Chiow}}, \bibinfo {author} {\bibfnamefont {S.}~\bibnamefont {Herrmann}}, \
  and\ \bibinfo {author} {\bibfnamefont {S.}~\bibnamefont {Chu}},\ }\href
  {\doibase 10.1103/PhysRevLett.102.240403} {\bibfield  {journal} {\bibinfo
  {journal} {Phys. Rev. Lett.}\ }\textbf {\bibinfo {volume} {102}},\ \bibinfo
  {pages} {240403} (\bibinfo {year} {2009})}\BibitemShut {NoStop}%
\bibitem [{\citenamefont {McDonald}\ \emph {et~al.}(2013)\citenamefont
  {McDonald}, \citenamefont {Kuhn}, \citenamefont {Bennetts}, \citenamefont
  {Debs}, \citenamefont {Hardman}, \citenamefont {Johnsson}, \citenamefont
  {Close},\ and\ \citenamefont {Robins}}]{McDonald2013}%
  \BibitemOpen
  \bibfield  {author} {\bibinfo {author} {\bibfnamefont {G.~D.}\ \bibnamefont
  {McDonald}}, \bibinfo {author} {\bibfnamefont {C.~C.~N.}\ \bibnamefont
  {Kuhn}}, \bibinfo {author} {\bibfnamefont {S.}~\bibnamefont {Bennetts}},
  \bibinfo {author} {\bibfnamefont {J.~E.}\ \bibnamefont {Debs}}, \bibinfo
  {author} {\bibfnamefont {K.~S.}\ \bibnamefont {Hardman}}, \bibinfo {author}
  {\bibfnamefont {M.}~\bibnamefont {Johnsson}}, \bibinfo {author}
  {\bibfnamefont {J.~D.}\ \bibnamefont {Close}}, \ and\ \bibinfo {author}
  {\bibfnamefont {N.~P.}\ \bibnamefont {Robins}},\ }\href {\doibase
  10.1103/PhysRevA.88.053620} {\bibfield  {journal} {\bibinfo  {journal} {Phys.
  Rev. A}\ }\textbf {\bibinfo {volume} {88}},\ \bibinfo {pages} {053620}
  (\bibinfo {year} {2013})}\BibitemShut {NoStop}%
\bibitem [{\citenamefont {Szigeti}\ \emph {et~al.}(2012)\citenamefont
  {Szigeti}, \citenamefont {Debs}, \citenamefont {Hope}, \citenamefont
  {Robins},\ and\ \citenamefont {Close}}]{Szigeti2012}%
  \BibitemOpen
  \bibfield  {author} {\bibinfo {author} {\bibfnamefont {S.~S.}\ \bibnamefont
  {Szigeti}}, \bibinfo {author} {\bibfnamefont {J.~E.}\ \bibnamefont {Debs}},
  \bibinfo {author} {\bibfnamefont {J.~J.}\ \bibnamefont {Hope}}, \bibinfo
  {author} {\bibfnamefont {N.~P.}\ \bibnamefont {Robins}}, \ and\ \bibinfo
  {author} {\bibfnamefont {J.~D.}\ \bibnamefont {Close}},\ }\href {\doibase
  10.1088/1367-2630/14/2/023009} {\bibfield  {journal} {\bibinfo  {journal}
  {New J. Phys.}\ }\textbf {\bibinfo {volume} {14}},\ \bibinfo {pages} {023009}
  (\bibinfo {year} {2012})}\BibitemShut {NoStop}%
\bibitem [{\citenamefont {Khaneja}\ \emph {et~al.}(2005)\citenamefont
  {Khaneja}, \citenamefont {Reiss}, \citenamefont {Kehlet}, \citenamefont
  {Schulte-Herbr{\"{u}}ggen},\ and\ \citenamefont {Glaser}}]{Khaneja2005}%
  \BibitemOpen
  \bibfield  {author} {\bibinfo {author} {\bibfnamefont {N.}~\bibnamefont
  {Khaneja}}, \bibinfo {author} {\bibfnamefont {T.}~\bibnamefont {Reiss}},
  \bibinfo {author} {\bibfnamefont {C.}~\bibnamefont {Kehlet}}, \bibinfo
  {author} {\bibfnamefont {T.}~\bibnamefont {Schulte-Herbr{\"{u}}ggen}}, \ and\
  \bibinfo {author} {\bibfnamefont {S.~J.}\ \bibnamefont {Glaser}},\ }\href
  {\doibase 10.1016/j.jmr.2004.11.004} {\bibfield  {journal} {\bibinfo
  {journal} {J. Magn. Reson.}\ }\textbf {\bibinfo {volume} {172}},\ \bibinfo
  {pages} {296} (\bibinfo {year} {2005})}\BibitemShut {NoStop}%
\bibitem [{\citenamefont {{De Fouquieres}}\ \emph {et~al.}(2011)\citenamefont
  {{De Fouquieres}}, \citenamefont {Schirmer}, \citenamefont {Glaser},\ and\
  \citenamefont {Kuprov}}]{DeFouquieres2011}%
  \BibitemOpen
  \bibfield  {author} {\bibinfo {author} {\bibfnamefont {P.}~\bibnamefont {{De
  Fouquieres}}}, \bibinfo {author} {\bibfnamefont {S.~G.}\ \bibnamefont
  {Schirmer}}, \bibinfo {author} {\bibfnamefont {S.~J.}\ \bibnamefont
  {Glaser}}, \ and\ \bibinfo {author} {\bibfnamefont {I.}~\bibnamefont
  {Kuprov}},\ }\href {\doibase 10.1016/j.jmr.2011.07.023} {\bibfield  {journal}
  {\bibinfo  {journal} {J. Magn. Reson.}\ }\textbf {\bibinfo {volume} {212}},\
  \bibinfo {pages} {412} (\bibinfo {year} {2011})}\BibitemShut {NoStop}%
\bibitem [{\citenamefont {Saywell}\ \emph {et~al.}(2018)\citenamefont
  {Saywell}, \citenamefont {Kuprov}, \citenamefont {Goodwin}, \citenamefont
  {Carey},\ and\ \citenamefont {Freegarde}}]{Saywell2018b}%
  \BibitemOpen
  \bibfield  {author} {\bibinfo {author} {\bibfnamefont {J.~C.}\ \bibnamefont
  {Saywell}}, \bibinfo {author} {\bibfnamefont {I.}~\bibnamefont {Kuprov}},
  \bibinfo {author} {\bibfnamefont {D.}~\bibnamefont {Goodwin}}, \bibinfo
  {author} {\bibfnamefont {M.}~\bibnamefont {Carey}}, \ and\ \bibinfo {author}
  {\bibfnamefont {T.}~\bibnamefont {Freegarde}},\ }\href {\doibase
  10.1103/PhysRevA.98.023625} {\bibfield  {journal} {\bibinfo  {journal} {Phys.
  Rev. A}\ }\textbf {\bibinfo {volume} {98}},\ \bibinfo {pages} {023625}
  (\bibinfo {year} {2018})}\BibitemShut {NoStop}%
\bibitem [{\citenamefont {Saywell}\ \emph {et~al.}(shed)\citenamefont
  {Saywell}, \citenamefont {Carey}, \citenamefont {Belal}, \citenamefont
  {Kuprov},\ and\ \citenamefont {Freegarde}}]{Saywell2020a}%
  \BibitemOpen
  \bibfield  {author} {\bibinfo {author} {\bibfnamefont {J.~C.}\ \bibnamefont
  {Saywell}}, \bibinfo {author} {\bibfnamefont {M.}~\bibnamefont {Carey}},
  \bibinfo {author} {\bibfnamefont {M.}~\bibnamefont {Belal}}, \bibinfo
  {author} {\bibfnamefont {I.}~\bibnamefont {Kuprov}}, \ and\ \bibinfo {author}
  {\bibfnamefont {T.}~\bibnamefont {Freegarde}},\ }\href
  {https://iopscience.iop.org/article/10.1088/1361-6455/ab6df6} {\bibfield
  {journal} {\bibinfo  {journal} {J. Phys. B}\ } (\bibinfo {year} {to be
  published})}\BibitemShut {NoStop}%
\bibitem [{\citenamefont {L{\'{e}}v{\`{e}}que}\ \emph
  {et~al.}(2009)\citenamefont {L{\'{e}}v{\`{e}}que}, \citenamefont {Gauguet},
  \citenamefont {Michaud}, \citenamefont {{Pereira Dos Santos}},\ and\
  \citenamefont {Landragin}}]{Leveque2009}%
  \BibitemOpen
  \bibfield  {author} {\bibinfo {author} {\bibfnamefont {T.}~\bibnamefont
  {L{\'{e}}v{\`{e}}que}}, \bibinfo {author} {\bibfnamefont {A.}~\bibnamefont
  {Gauguet}}, \bibinfo {author} {\bibfnamefont {F.}~\bibnamefont {Michaud}},
  \bibinfo {author} {\bibfnamefont {F.}~\bibnamefont {{Pereira Dos Santos}}}, \
  and\ \bibinfo {author} {\bibfnamefont {A.}~\bibnamefont {Landragin}},\ }\href
  {\doibase 10.1103/PhysRevLett.103.080405} {\bibfield  {journal} {\bibinfo
  {journal} {Phys. Rev. Lett.}\ }\textbf {\bibinfo {volume} {103}},\ \bibinfo
  {pages} {080405} (\bibinfo {year} {2009})}\BibitemShut {NoStop}%
\bibitem [{\citenamefont {Malossi}\ \emph {et~al.}(2010)\citenamefont
  {Malossi}, \citenamefont {Bodart}, \citenamefont {Merlet}, \citenamefont
  {L{\'{e}}v{\`{e}}que}, \citenamefont {Landragin},\ and\ \citenamefont
  {Pereira Dos~Santos}}]{Malossi2010a}%
  \BibitemOpen
  \bibfield  {author} {\bibinfo {author} {\bibfnamefont {N.}~\bibnamefont
  {Malossi}}, \bibinfo {author} {\bibfnamefont {Q.}~\bibnamefont {Bodart}},
  \bibinfo {author} {\bibfnamefont {S.}~\bibnamefont {Merlet}}, \bibinfo
  {author} {\bibfnamefont {T.}~\bibnamefont {L{\'{e}}v{\`{e}}que}}, \bibinfo
  {author} {\bibfnamefont {A.}~\bibnamefont {Landragin}}, \ and\ \bibinfo
  {author} {\bibfnamefont {F.}~\bibnamefont {Pereira Dos~Santos}},\ }\href
  {\doibase 10.1103/PhysRevA.81.013617} {\bibfield  {journal} {\bibinfo
  {journal} {Phys. Rev. A}\ }\textbf {\bibinfo {volume} {81}},\ \bibinfo
  {pages} {013617} (\bibinfo {year} {2010})}\BibitemShut {NoStop}%
\bibitem [{\citenamefont {Kotru}(2016)}]{Kotru2016a}%
  \BibitemOpen
  \bibfield  {author} {\bibinfo {author} {\bibfnamefont {K.}~\bibnamefont
  {Kotru}},\ }\emph {\bibinfo {title} {{Timekeeping and Accelerometry with
  Robust Light Pulse Atom Interferometers}}},\ \href@noop {} {Ph.D. thesis},\
  \bibinfo  {school} {Massachusetts Institute of Technology} (\bibinfo {year}
  {2016})\BibitemShut {NoStop}%
\bibitem [{\citenamefont {Jaffe}(2019)}]{Jaffe2019}%
  \BibitemOpen
  \bibfield  {author} {\bibinfo {author} {\bibfnamefont {M.}~\bibnamefont
  {Jaffe}},\ }\emph {\bibinfo {title} {{Atom Interferometry in an Optical
  Cavity}}},\ \href@noop {} {Ph.D. thesis},\ \bibinfo  {school} {University of
  California, Berkeley} (\bibinfo {year} {2019})\BibitemShut {NoStop}%
\bibitem [{\citenamefont {Shaka}\ \emph {et~al.}(1983)\citenamefont {Shaka},
  \citenamefont {Keeler}, \citenamefont {Frenkiel},\ and\ \citenamefont
  {Freeman}}]{Shaka1983}%
  \BibitemOpen
  \bibfield  {author} {\bibinfo {author} {\bibfnamefont {A.~J.}\ \bibnamefont
  {Shaka}}, \bibinfo {author} {\bibfnamefont {J.}~\bibnamefont {Keeler}},
  \bibinfo {author} {\bibfnamefont {T.}~\bibnamefont {Frenkiel}}, \ and\
  \bibinfo {author} {\bibfnamefont {R.}~\bibnamefont {Freeman}},\ }\href
  {\doibase 10.1016/0022-2364(83)90207-X} {\bibfield  {journal} {\bibinfo
  {journal} {J. Magn. Reson. (1969)}\ }\textbf {\bibinfo {volume} {52}},\
  \bibinfo {pages} {335} (\bibinfo {year} {1983})}\BibitemShut {NoStop}%
\bibitem [{\citenamefont {Kovachy}\ \emph {et~al.}(2012)\citenamefont
  {Kovachy}, \citenamefont {Chiow},\ and\ \citenamefont
  {Kasevich}}]{Kovachy2012a}%
  \BibitemOpen
  \bibfield  {author} {\bibinfo {author} {\bibfnamefont {T.}~\bibnamefont
  {Kovachy}}, \bibinfo {author} {\bibfnamefont {S.-w.}\ \bibnamefont {Chiow}},
  \ and\ \bibinfo {author} {\bibfnamefont {M.~A.}\ \bibnamefont {Kasevich}},\
  }\href {\doibase 10.1103/PhysRevA.86.011606} {\bibfield  {journal} {\bibinfo
  {journal} {Phys. Rev. A}\ }\textbf {\bibinfo {volume} {86}},\ \bibinfo
  {pages} {011606(R)} (\bibinfo {year} {2012})}\BibitemShut {NoStop}%
\bibitem [{\citenamefont {Jaffe}\ \emph {et~al.}(2018)\citenamefont {Jaffe},
  \citenamefont {Xu}, \citenamefont {Haslinger}, \citenamefont {M{\"{u}}ller},\
  and\ \citenamefont {Hamilton}}]{Jaffe2018}%
  \BibitemOpen
  \bibfield  {author} {\bibinfo {author} {\bibfnamefont {M.}~\bibnamefont
  {Jaffe}}, \bibinfo {author} {\bibfnamefont {V.}~\bibnamefont {Xu}}, \bibinfo
  {author} {\bibfnamefont {P.}~\bibnamefont {Haslinger}}, \bibinfo {author}
  {\bibfnamefont {H.}~\bibnamefont {M{\"{u}}ller}}, \ and\ \bibinfo {author}
  {\bibfnamefont {P.}~\bibnamefont {Hamilton}},\ }\href {\doibase
  10.1103/PhysRevLett.121.040402} {\bibfield  {journal} {\bibinfo  {journal}
  {Phys. Rev. Lett.}\ }\textbf {\bibinfo {volume} {121}},\ \bibinfo {pages}
  {040402} (\bibinfo {year} {2018})}\BibitemShut {NoStop}%
\bibitem [{\citenamefont {Feynman}\ \emph {et~al.}(1957)\citenamefont
  {Feynman}, \citenamefont {Vernon},\ and\ \citenamefont
  {Hellwarth}}]{Feynman1957}%
  \BibitemOpen
  \bibfield  {author} {\bibinfo {author} {\bibfnamefont {R.~P.}\ \bibnamefont
  {Feynman}}, \bibinfo {author} {\bibfnamefont {F.~L.}\ \bibnamefont {Vernon}},
  \ and\ \bibinfo {author} {\bibfnamefont {R.~W.}\ \bibnamefont {Hellwarth}},\
  }\href {\doibase 10.1063/1.1722572} {\bibfield  {journal} {\bibinfo
  {journal} {J. Appl. Phys.}\ }\textbf {\bibinfo {volume} {28}},\ \bibinfo
  {pages} {49} (\bibinfo {year} {1957})}\BibitemShut {NoStop}%
\bibitem [{\citenamefont {Hardy}\ \emph {et~al.}(1986)\citenamefont {Hardy},
  \citenamefont {Edelstein},\ and\ \citenamefont {Vatis}}]{Hardy1986}%
  \BibitemOpen
  \bibfield  {author} {\bibinfo {author} {\bibfnamefont {C.}~\bibnamefont
  {Hardy}}, \bibinfo {author} {\bibfnamefont {W.}~\bibnamefont {Edelstein}}, \
  and\ \bibinfo {author} {\bibfnamefont {D.}~\bibnamefont {Vatis}},\ }\href
  {\doibase 10.1016/0022-2364(86)90190-3} {\bibfield  {journal} {\bibinfo
  {journal} {J. Magn. Reson. (1969)}\ }\textbf {\bibinfo {volume} {66}},\
  \bibinfo {pages} {470} (\bibinfo {year} {1986})}\BibitemShut {NoStop}%
\bibitem [{\citenamefont {Baum}\ \emph {et~al.}(1985)\citenamefont {Baum},
  \citenamefont {Tycko},\ and\ \citenamefont {Pines}}]{Baum1985b}%
  \BibitemOpen
  \bibfield  {author} {\bibinfo {author} {\bibfnamefont {J.}~\bibnamefont
  {Baum}}, \bibinfo {author} {\bibfnamefont {R.}~\bibnamefont {Tycko}}, \ and\
  \bibinfo {author} {\bibfnamefont {A.}~\bibnamefont {Pines}},\ }\href
  {\doibase 10.1103/PhysRevA.32.3435} {\bibfield  {journal} {\bibinfo
  {journal} {Phys. Rev. A}\ }\textbf {\bibinfo {volume} {32}},\ \bibinfo
  {pages} {3435} (\bibinfo {year} {1985})}\BibitemShut {NoStop}%
\bibitem [{\citenamefont {Tann{\'{u}}s}\ and\ \citenamefont
  {Garwood}(1996)}]{Tannus1996}%
  \BibitemOpen
  \bibfield  {author} {\bibinfo {author} {\bibfnamefont {A.}~\bibnamefont
  {Tann{\'{u}}s}}\ and\ \bibinfo {author} {\bibfnamefont {M.}~\bibnamefont
  {Garwood}},\ }\href {\doibase 10.1006/jmra.1996.0110} {\bibfield  {journal}
  {\bibinfo  {journal} {J. Magn. Reson., Series A}\ }\textbf {\bibinfo {volume}
  {120}},\ \bibinfo {pages} {133} (\bibinfo {year} {1996})}\BibitemShut
  {NoStop}%
\bibitem [{\citenamefont {Hwang}\ \emph {et~al.}(1998)\citenamefont {Hwang},
  \citenamefont {van Zijl},\ and\ \citenamefont {Garwood}}]{Hwang1998}%
  \BibitemOpen
  \bibfield  {author} {\bibinfo {author} {\bibfnamefont {T.-L.}\ \bibnamefont
  {Hwang}}, \bibinfo {author} {\bibfnamefont {P.~C.}\ \bibnamefont {van Zijl}},
  \ and\ \bibinfo {author} {\bibfnamefont {M.}~\bibnamefont {Garwood}},\ }\href
  {\doibase 10.1006/jmre.1998.1441} {\bibfield  {journal} {\bibinfo  {journal}
  {J. Magn. Reson.}\ }\textbf {\bibinfo {volume} {133}},\ \bibinfo {pages}
  {200} (\bibinfo {year} {1998})}\BibitemShut {NoStop}%
\bibitem [{\citenamefont {Garwood}\ and\ \citenamefont
  {DelaBarre}(2001)}]{Garwood2001a}%
  \BibitemOpen
  \bibfield  {author} {\bibinfo {author} {\bibfnamefont {M.}~\bibnamefont
  {Garwood}}\ and\ \bibinfo {author} {\bibfnamefont {L.}~\bibnamefont
  {DelaBarre}},\ }\href {\doibase 10.1006/jmre.2001.2340} {\bibfield  {journal}
  {\bibinfo  {journal} {J. Magn. Reson.}\ }\textbf {\bibinfo {volume} {153}},\
  \bibinfo {pages} {155} (\bibinfo {year} {2001})}\BibitemShut {NoStop}%
\bibitem [{\citenamefont {Kobzar}\ \emph {et~al.}(2004)\citenamefont {Kobzar},
  \citenamefont {Skinner}, \citenamefont {Khaneja}, \citenamefont {Glaser},\
  and\ \citenamefont {Luy}}]{Kobzar2004}%
  \BibitemOpen
  \bibfield  {author} {\bibinfo {author} {\bibfnamefont {K.}~\bibnamefont
  {Kobzar}}, \bibinfo {author} {\bibfnamefont {T.~E.}\ \bibnamefont {Skinner}},
  \bibinfo {author} {\bibfnamefont {N.}~\bibnamefont {Khaneja}}, \bibinfo
  {author} {\bibfnamefont {S.~J.}\ \bibnamefont {Glaser}}, \ and\ \bibinfo
  {author} {\bibfnamefont {B.}~\bibnamefont {Luy}},\ }\href {\doibase
  10.1016/j.jmr.2004.06.017} {\bibfield  {journal} {\bibinfo  {journal} {J.
  Magn. Reson.}\ }\textbf {\bibinfo {volume} {170}},\ \bibinfo {pages} {236}
  (\bibinfo {year} {2004})}\BibitemShut {NoStop}%
\bibitem [{\citenamefont {Kobzar}\ \emph {et~al.}(2012)\citenamefont {Kobzar},
  \citenamefont {Ehni}, \citenamefont {Skinner}, \citenamefont {Glaser},\ and\
  \citenamefont {Luy}}]{Kobzar2012}%
  \BibitemOpen
  \bibfield  {author} {\bibinfo {author} {\bibfnamefont {K.}~\bibnamefont
  {Kobzar}}, \bibinfo {author} {\bibfnamefont {S.}~\bibnamefont {Ehni}},
  \bibinfo {author} {\bibfnamefont {T.~E.}\ \bibnamefont {Skinner}}, \bibinfo
  {author} {\bibfnamefont {S.~J.}\ \bibnamefont {Glaser}}, \ and\ \bibinfo
  {author} {\bibfnamefont {B.}~\bibnamefont {Luy}},\ }\href {\doibase
  10.1016/j.jmr.2012.09.013} {\bibfield  {journal} {\bibinfo  {journal} {J.
  Magn. Reson.}\ }\textbf {\bibinfo {volume} {225}},\ \bibinfo {pages} {142}
  (\bibinfo {year} {2012})}\BibitemShut {NoStop}%
\bibitem [{\citenamefont {Stoner}\ \emph {et~al.}(2011)\citenamefont {Stoner},
  \citenamefont {Butts}, \citenamefont {Kinast},\ and\ \citenamefont
  {Timmons}}]{Stoner2011}%
  \BibitemOpen
  \bibfield  {author} {\bibinfo {author} {\bibfnamefont {R.}~\bibnamefont
  {Stoner}}, \bibinfo {author} {\bibfnamefont {D.}~\bibnamefont {Butts}},
  \bibinfo {author} {\bibfnamefont {J.}~\bibnamefont {Kinast}}, \ and\ \bibinfo
  {author} {\bibfnamefont {B.}~\bibnamefont {Timmons}},\ }\href {\doibase
  10.1364/JOSAB.28.002418} {\bibfield  {journal} {\bibinfo  {journal} {J. Opt.
  Soc. Am. B}\ }\textbf {\bibinfo {volume} {28}},\ \bibinfo {pages} {2418}
  (\bibinfo {year} {2011})}\BibitemShut {NoStop}%
\bibitem [{\citenamefont {Kobzar}\ \emph {et~al.}(2005)\citenamefont {Kobzar},
  \citenamefont {Luy}, \citenamefont {Khaneja},\ and\ \citenamefont
  {Glaser}}]{Kobzar2005}%
  \BibitemOpen
  \bibfield  {author} {\bibinfo {author} {\bibfnamefont {K.}~\bibnamefont
  {Kobzar}}, \bibinfo {author} {\bibfnamefont {B.}~\bibnamefont {Luy}},
  \bibinfo {author} {\bibfnamefont {N.}~\bibnamefont {Khaneja}}, \ and\
  \bibinfo {author} {\bibfnamefont {S.~J.}\ \bibnamefont {Glaser}},\ }\href
  {\doibase 10.1016/j.jmr.2004.12.005} {\bibfield  {journal} {\bibinfo
  {journal} {J. Magn. Reson.}\ }\textbf {\bibinfo {volume} {173}},\ \bibinfo
  {pages} {229} (\bibinfo {year} {2005})}\BibitemShut {NoStop}%
\bibitem [{\citenamefont {Braun}\ and\ \citenamefont
  {Glaser}(2014)}]{Braun2014}%
  \BibitemOpen
  \bibfield  {author} {\bibinfo {author} {\bibfnamefont {M.}~\bibnamefont
  {Braun}}\ and\ \bibinfo {author} {\bibfnamefont {S.~J.}\ \bibnamefont
  {Glaser}},\ }\href {\doibase 10.1088/1367-2630/16/11/115002} {\bibfield
  {journal} {\bibinfo  {journal} {New J. Phys.}\ }\textbf {\bibinfo {volume}
  {16}},\ \bibinfo {pages} {115002} (\bibinfo {year} {2014})}\BibitemShut
  {NoStop}%
\bibitem [{\citenamefont {Emsley}\ \emph {et~al.}(1990)\citenamefont {Emsley},
  \citenamefont {Burghardt},\ and\ \citenamefont {Bodenhausen}}]{Emsley1990}%
  \BibitemOpen
  \bibfield  {author} {\bibinfo {author} {\bibfnamefont {L.}~\bibnamefont
  {Emsley}}, \bibinfo {author} {\bibfnamefont {I.}~\bibnamefont {Burghardt}}, \
  and\ \bibinfo {author} {\bibfnamefont {G.}~\bibnamefont {Bodenhausen}},\
  }\href {\doibase 10.1016/0022-2364(90)90381-I} {\bibfield  {journal}
  {\bibinfo  {journal} {J. Magn. Reson. (1969)}\ }\textbf {\bibinfo {volume}
  {90}},\ \bibinfo {pages} {214} (\bibinfo {year} {1990})}\BibitemShut
  {NoStop}%
\bibitem [{\citenamefont {Geen}\ and\ \citenamefont
  {Freeman}(1991)}]{Geen1991}%
  \BibitemOpen
  \bibfield  {author} {\bibinfo {author} {\bibfnamefont {H.}~\bibnamefont
  {Geen}}\ and\ \bibinfo {author} {\bibfnamefont {R.}~\bibnamefont {Freeman}},\
  }\href {\doibase 10.1016/0022-2364(91)90034-Q} {\bibfield  {journal}
  {\bibinfo  {journal} {J. Magn. Reson. (1969)}\ }\textbf {\bibinfo {volume}
  {93}},\ \bibinfo {pages} {93} (\bibinfo {year} {1991})}\BibitemShut {NoStop}%
\bibitem [{\citenamefont {Freeman}(1998)}]{Freeman1998a}%
  \BibitemOpen
  \bibfield  {author} {\bibinfo {author} {\bibfnamefont {R.}~\bibnamefont
  {Freeman}},\ }\href {\doibase 10.1016/S0079-6565(97)00024-1} {\bibfield
  {journal} {\bibinfo  {journal} {Prog. Nucl. Mag. Res. Spectrosc.}\ }\textbf
  {\bibinfo {volume} {32}},\ \bibinfo {pages} {59} (\bibinfo {year}
  {1998})}\BibitemShut {NoStop}%
\bibitem [{\citenamefont {Janich}\ \emph {et~al.}(2011)\citenamefont {Janich},
  \citenamefont {Schulte}, \citenamefont {Schwaiger},\ and\ \citenamefont
  {Glaser}}]{Janich2011}%
  \BibitemOpen
  \bibfield  {author} {\bibinfo {author} {\bibfnamefont {M.~A.}\ \bibnamefont
  {Janich}}, \bibinfo {author} {\bibfnamefont {R.~F.}\ \bibnamefont {Schulte}},
  \bibinfo {author} {\bibfnamefont {M.}~\bibnamefont {Schwaiger}}, \ and\
  \bibinfo {author} {\bibfnamefont {S.~J.}\ \bibnamefont {Glaser}},\ }\href
  {\doibase 10.1016/j.jmr.2011.09.025} {\bibfield  {journal} {\bibinfo
  {journal} {J. Magn. Reson.}\ }\textbf {\bibinfo {volume} {213}},\ \bibinfo
  {pages} {126} (\bibinfo {year} {2011})}\BibitemShut {NoStop}%
\bibitem [{\citenamefont {Warren}(1984)}]{Warren1984}%
  \BibitemOpen
  \bibfield  {author} {\bibinfo {author} {\bibfnamefont {W.~S.}\ \bibnamefont
  {Warren}},\ }\href {\doibase 10.1063/1.447644} {\bibfield  {journal}
  {\bibinfo  {journal} {J. Chem. Phys.}\ }\textbf {\bibinfo {volume} {81}},\
  \bibinfo {pages} {5437} (\bibinfo {year} {1984})}\BibitemShut {NoStop}%
\bibitem [{\citenamefont {Hogben}\ \emph {et~al.}(2011)\citenamefont {Hogben},
  \citenamefont {Krzystyniak}, \citenamefont {Charnock}, \citenamefont {Hore},\
  and\ \citenamefont {Kuprov}}]{Hogben2011}%
  \BibitemOpen
  \bibfield  {author} {\bibinfo {author} {\bibfnamefont {H.~J.}\ \bibnamefont
  {Hogben}}, \bibinfo {author} {\bibfnamefont {M.}~\bibnamefont {Krzystyniak}},
  \bibinfo {author} {\bibfnamefont {G.~T.~P.}\ \bibnamefont {Charnock}},
  \bibinfo {author} {\bibfnamefont {P.~J.}\ \bibnamefont {Hore}}, \ and\
  \bibinfo {author} {\bibfnamefont {I.}~\bibnamefont {Kuprov}},\ }\href
  {\doibase 10.1016/j.jmr.2010.11.008} {\bibfield  {journal} {\bibinfo
  {journal} {J. Magn. Reson.}\ }\textbf {\bibinfo {volume} {208}},\ \bibinfo
  {pages} {179} (\bibinfo {year} {2011})}\BibitemShut {NoStop}%
\bibitem [{\citenamefont {Goodwin}\ and\ \citenamefont
  {Kuprov}(2016)}]{Goodwin2015}%
  \BibitemOpen
  \bibfield  {author} {\bibinfo {author} {\bibfnamefont {D.~L.}\ \bibnamefont
  {Goodwin}}\ and\ \bibinfo {author} {\bibfnamefont {I.}~\bibnamefont
  {Kuprov}},\ }\href {\doibase 10.1063/1.4949534} {\bibfield  {journal}
  {\bibinfo  {journal} {J. Chem. Phys.}\ }\textbf {\bibinfo {volume} {144}},\
  \bibinfo {pages} {204107} (\bibinfo {year} {2016})}\BibitemShut {NoStop}%
\end{thebibliography}%

\end{document}